\documentclass{article}

 \usepackage{
 	amsmath,amssymb,amstext,amsthm,  
 	graphicx,    
 	mathtools,   
 	bbm,         
 	soul,         
 	subfigure, float, enumitem, lscape, 
 	ifthen, 
 	comment      
 }
 \usepackage{
 	colortbl, pdflscape, booktabs, longtable, array, wrapfig, float, tabu,afterpage, lipsum, threeparttable, multirow, 
 	ragged2e, bm, wrapfig, booktabs, pdfpages,
 	chngcntr  
 }  
 \usepackage[ruled,vlined]{algorithm2e}
 \usepackage[affil-it]{authblk} 
 \usepackage{setspace} 
 \newboolean{PrintVersion}
 \setboolean{PrintVersion}{true}

\usepackage[pdftex,pagebackref=false]{hyperref} 

\newcommand{\BibPath}{.}
\usepackage[style=apa, sortcites = true, url=true,  backend=biber,natbib=true]{biblatex} 
\DeclareFieldFormat{apacase}{#1}  
\usepackage{tikz}
\usetikzlibrary{shapes, arrows} 
\usepackage{pgfplots}
\pgfplotsset{compat=1.10}
\usepgfplotslibrary{fillbetween}

\ifthenelse{\boolean{PrintVersion}}{   
\hypersetup{	
    colorlinks,%
    citecolor=black,%
    filecolor=black,%
    linkcolor=black,%
    urlcolor=black}
}{} 

\let\origdoublepage\cleardoublepage
\newcommand{\clearemptydoublepage}{%
  \clearpage{\pagestyle{empty}\origdoublepage}}
\let\cleardoublepage\clearemptydoublepage

\newtheorem{theorem}{Theorem}[section]
   
\newtheorem{prop}[theorem]{Proposition}        
\newtheorem{remark}[theorem]{Remark}           
\newtheorem{example}[theorem]{Example}        
\newtheorem{coro}[theorem]{Corollary}         
        
\newtheorem{lemma}[theorem]{Lemma}

\definecolor{mayablue}{rgb}{0.45, 0.76, 0.98}

\definecolor{lightblue}{rgb}{0.68, 0.85, 0.9}

\def\E{\mathrm{E}}

\def\Var{\mathrm{Var}}

\newcommand*\diff{\mathop{}\!\mathrm{d}}
\DeclareMathOperator*{\argmin}{\arg\!\min}

\title{Option Pricing with Time-Changed Fractional Brownian Motion: A Fractional Variance Gamma Model}

\author[1]{Robert Jarrow\thanks{email: robertjarrow@cornell.edu}}
\author[1,2]{Jayen Tan\thanks{email: jayen.tan@ntu.edu.sg}}
\affil[1]{\small Samuel Curtis Johnson Graduate School of Management, Cornell University} 
\affil[2]{\small Division of Finance, Nanyang Business School, Nanyang Technological University}

\begin{document}

\maketitle 

\begin{abstract} 
	Fractional Brownian motion (fBm) exhibits attractive features for financial modeling, including long-range dependence, path roughness, and anomalous diffusion. However, its non-semimartingale nature precludes the use of conventional no-arbitrage approaches to option pricing. We address this limitation by introducing a time-changed fBm, obtained by evaluating fBm at stochastic gamma activity time, where activity time represents cumulative executed trading time. The resulting process retains the defining properties of fBm while recovering the semimartingale structure. Building on this construction, we develop the fractional Variance Gamma (fVG) model and propose a generalized method of moments (GMM) estimation procedure for option pricing. An empirical analysis of the S\&P 500 yields an estimated Hurst exponent of approximately 0.45, consistent with mildly sublinear temporal scaling of return moments.
	
	
\end{abstract}

\section{Introduction} 

While Brownian motion remains the canonical continuous-time stochastic process for modeling financial time series, it does not capture several empirical regularities documented in financial markets. Fractional Brownian motion (fBm) generalizes classical Brownian motion by accommodating long-range dependence, path roughness, and anomalous diffusion, allowing fBm-driven processes to incorporate empirical features that are absent from the classical Brownian setting.

One important application is option pricing, where stock prices serve as the underlying asset. Several empirical studies report evidence of long-range dependence in stock returns, challenging the short-memory assumption embedded in standard diffusion models \citep{Willinger1999LRDFinance, GreeneFielitz1977LRDStockReturns, Christodoulou2006LRDStocks}. In addition, a growing body of evidence documents multifractal scaling in asset returns across temporal horizons \citep[e.g.,][]{MandelbrotEtAl1997MultifractalModelAsset, CalvetFisher2002MultifractalityAssetReturns, Caraiani2012EvidenceMultifractalityEmerging, DiMatteo2007MultiscalingFinance, SuarezGarciaGomezUllate2014MultifractalityLongMemory, TurielPerezVicente2003MultifractalGeometryStock}. For instance,  \citet{MantegnaStanley1995ScalingBehaviourDynamics} show that the distribution of S\&P 500 returns measured over different time intervals satisfies a power-law scaling relation with a Lévy exponent of approximately 1.4.
Complementing this evidence, \citet{LoMacKinlay1988StockPricesNotRandomWalk} reject the random walk hypothesis using the variance ratio test, showing that the variance ratio increases more than proportionally with the sampling horizon. Similar findings have subsequently been reported for Asian \citep{CharlesDarne2009VarianceRatioTestsRandom, HoqueEtAl2007ComparisonVarianceRatio, RyooSmith2002KoreanStockVarianceRatio}, European \citep{SmithRyoo2003EuroVarianceRatioTests}, and African equity markets \citep{SmithEtAl2002AfricanVarianceRatio}; see also \citet{LoMacKinlay1989SizePowerVarianceRatioTest, CharlesDarne2009VarianceRatioTestsRandom}.


However, fBm is not a (semi)martingale, which presents a fundamental obstacle when applying the standard continuous-time,
continuous-trading, arbitrage-free pricing framework.\footnote{See \citet[Chapter 2]{Jarrow2021CtsTimeAssetPricing} for a textbook presentation of the standard arbitrage-free pricing model.}
Specifically, the fundamental theorems of asset pricing rely critically on the semimartingale property of the underlying asset. 
When price process is not a semimartingale, Itô calculus no longer applies \citep{Protter2005TBChapSemimartingales}, and arbitrage opportunities inevitably arise in the standard frictionless setting \citep{DelbaenSchachermayer1994FundamentalTheorems}.
Consistent with this result, \citet{Rogers1997ArbitragefBm} and \citet{Cheridito2003ArbitragefBm} show that the fractional Black--Scholes--Merton model admits arbitrage under continuous trading. Section~\ref{sec: Literature Review on Fractional Asset Pricing Models} provides a comprehensive review of fractional stock price models.

In this paper, we propose a novel approach to option pricing by introducing a time-changed fBm process, defined as $X=B_H(\gamma)$, where the fBm $B_H$ is evaluated at stochastic gamma activity time $\gamma$. The use of activity time has a long tradition in option pricing; see e.g., \citet{MadanCarrChang1998VGoption, GemanEtAl2001TimeChangesLevy, MadanWang2023EconomicsTime}.
Economically, activity time represents the accumulation of trading activity and the arrival of market information. Modeling asset dynamics in activity time rather than calendar time therefore provides a natural way to account for variations in trading intensity. In this view, price movements are driven by the arrival of new information through trading activity rather than solely by the passage of calendar time.

The time-changed fBm preserves the principal statistical properties of standard fBm while remaining analytically tractable. In particular, its distributional properties can be characterized through a Gaussian--Gamma mixture representation. More importantly, our central theoretical result establishes that the time-changed fBm is a semimartingale, rendering it compatible with the classical no-arbitrage pricing framework based on equivalent martingale measures. Building on this result, we introduce the fractional Variance Gamma (fVG) model, which extends the Variance Gamma (VG) model of \citet{MadanSeneta1990VGmodel} by replacing the Brownian motion component with fBm. The resulting five-parameter specification captures the drift, volatility, skewness, kurtosis, and dependence structure of asset returns. Unlike the classical VG model, however, the fVG process does not possess independent increments and does not admit a closed-form density. Consequently, standard pricing and estimation methods that rely on independent increments or closed-form expressions are no longer directly applicable. To address these challenges, we develop a marked point process representation of the time-changed fBm, characterize its compensator and derive closed-form expressions for its raw and central moments.

Under the fVG model, we show that stock returns admit a decomposition into four distinct components: (i) compensation for the risk-free rate, (ii) a risk-premium, (iii) an endogenous drift generated by the dependence structure and historical trajectory of the process, and (iv) a martingale innovation that captures residual uncertainty. 
Unlike classical semimartingale models, the fVG stock price specification therefore contains a dependence-induced drift through which the current state of the process influences its future evolution. This provides an economically interpretable channel through which persistence in returns affects asset prices.
We further consider an alternative stock price specification based on time-changed fractional geometric Brownian motion and establish its equivalence to the proposed fVG model.

To compute the model's conditional expectations, we construct a simulation procedure that first generates the discretized sample paths of the fBm and gamma processes independently, and then evaluates the fBm at the realized gamma activity times. To ensure numerical accuracy, the fBm is generated on a finer (secondary) grid than that used for the gamma activity time (primary grid).

For parameter estimation, we propose a generalized method of moments (GMM) procedure for the fVG model that exploits the unconditional moments of stock returns across multiple time horizons. To illustrate the proposed procedure, we fit the fVG model to S\&P 500 index and compare the full specification with several nested benchmark models, including the classical Black--Scholes--Merton (BSM), Variance Gamma (VG), and fractional Black--Scholes--Merton (fBSM) models.
Strikingly, the estimated Hurst parameter $H$ under the full fVG specification is approximately 0.45, which lies below the Brownian benchmark of 0.5.

This result contrasts with studies reporting long range dependence in asset returns. It suggests that some of the apparent long memory documented in simpler models may partially reflect unmodeled heavy-tailed behavior rather than genuine temporal dependence. Our GMM estimator identifies $H$ primarily through the scaling behavior of return moments across investment horizons, that is, through the rate at which these moments accumulate with lag length. Consequently, the estimate $H \approx 0.45$ implies that the term structure of empirical moments exhibits mildly sub-linear temporal scaling.

Overall, the proposed fVG model accommodates fractional dynamics in stock price evolution while preserving compatibility with the classical arbitrage-free pricing paradigm. More broadly, the underlying time-changed fBm provides a flexible stochastic process for a broad range of financial time series applications. Although time-changed fBm has appeared in the statistical literature, to the best of our knowledge, this is the first paper to incorporate such a process into an arbitrage-free asset-pricing framework. We leave a comprehensive empirical evaluation of the fVG option pricing model to future research.

The remainder of the paper is organized as follows. 
Section~\ref{sec: Literature Review on Fractional Asset Pricing Models} reviews the related literature. 
Section~\ref{sec: Stochastic Processes} provides background on the stochastic processes underlying our analysis and establishes key results used throughout the subsequent analysis. 
Section~\ref{sec: Time-changed fBm process} introduces the time-changed fBm. 
Section~\ref{sec: fVG model} develops the fVG equity model. Section~\ref{sec: Parameter Estimation} presents a feasible GMM procedure for parameter estimation. 
Section~\ref{sec: Application to SP500} reports an empirical application to the S\&P 500 index. 
Section~\ref{sec: Conclusion} concludes. 
Proofs of the main theorems and propositions are collected in Appendix~\ref{sec: Proofs}.

\section{Literature Review on Fractional Asset Pricing Models}
\label{sec: Literature Review on Fractional Asset Pricing Models}

Modern continuous-time, continuous-trading option pricing theory is built on the assumption that asset prices follow semimartingale processes. This assumption is fundamental because semimartingales constitute the largest class of stochastic processes for which Itô calculus is well defined (e.g., see \citet{Protter2005TBChapSemimartingales}), providing the mathematical foundation for much of continuous-time financial economics.
Although fBm possesses several empirically appealing properties --- including long-range dependence, path roughness, and anomalous diffusion --- it is not a semimartingale. Over the years, several approaches have been developed to incorporate fBm into financial modeling.


The earliest attempts adopted a direct substitution approach, replacing Brownian motion with fBm in the classical Black--Scholes--Merton (BSM) framework to obtain a fractional BSM model. However, this formulation was shown to be incompatible with the no-arbitrage framework of continuous-time finance.
Under the standard continuous-time, continuous-trading setting, the fundamental results of \citet{DelbaenSchachermayer1994FundamentalTheorems} establish that arbitrage-free pricing requires the underlying price process to be a semimartingale. Consistent with this result, several studies have constructed explicit arbitrage strategies for fBm-driven asset prices, demonstrating that the fractional BSM model admits arbitrage under continuous trading \citep{Rogers1997ArbitragefBm, Cheridito2003ArbitragefBm}.

To circumvent the arbitrage issues arising from the non-semimartingale nature of fBm, a second approach modifies the definitions of arbitrage and self-financing trading strategies to develop an analytically tractable pricing framework based on divergence/Wick calculus \citep{HuOksendal2003fractionalWhiteNoise, ElliottVan2003fractionalWhiteNoise}. However, these modified concepts fail to preserve the standard economic interpretation of portfolio values and trading strategies, limiting their applicability and relevance in financial valuation \citep{BjorkHult2005WickProduct}.

Separately, an alternative approach in the literature incorporates fBm into asset pricing models by modifying the frictionless market assumption. The central idea is to restrict the class of admissible self-financing trading strategies to restore the absence of free-lunch-with-vanishing-risk (FLVR). For example, \citet{Cheridito2003ArbitragefBm} and \citet{JarrowProtterSayit2009NoArbitrageSemimartingales} consider trading strategies that impose a strictly positive minimum time interval between consecutive trades. \citet{BenderEtAl2008PricingNonSemimartingale} restrict admissible strategies to those representable as Borel-measurable functions of a finite, prespecified collection of adapted functionals of the stock price path. Alternatively, \citet{Guasoni2006fBmTransactionCost} introduces proportional transaction costs and restricts trading strategies to processes of finite variation. While these approaches successfully restore the FLVR condition, they do so by departing from the classical frictionless continuous-trading framework. Consequently, the Fundamental Theorems of Asset Pricing no longer apply directly, and option valuation cannot generally be formulated through equivalent martingale measures. This departure limits the analytical tractability of these approaches.

Another branch of the literature constructs alternative semimartingale processes that replicates selected features of fBm, such as long-range dependence and self-similarity. Examples include time-changed Brownian motion with long-range dependent activity times \citep{CalvetFisher2002MultifractalityAssetReturns,HeydeLeonenko2005StudentActivityTimeLRD,FinlaySeneta2006ActivityTimeLRD,FinlaySeneta2007GammaActivityTimeLRD,FinlaySenetaWang2012InverseGammaActivityTime} and regularizations of fBm \citep{Cheridito2001MixedfBm,Rogers1997ArbitragefBm}. While these approaches retain compatibility with semimartingale-based asset pricing, they typically introduce additional modeling complexity and depart from the canonical Brownian-motion-based paradigm.

Finally, \citet{TanJarrow2025ArbitragefreefBm} develop an arbitrage-free valuation framework for derivatives written on non-traded fractional processes. Their approach establishes a coherent pricing and replication theory for derivatives written on a non-semimartingale underlying, provided that the underlying process itself is excluded from direct trading. Arbitrage-free valuation is achieved through trading in a sufficiently rich set of derivatives written on the underlying, allowing the standard arbitrage-free pricing methodology to be applied despite the non-semimartingale nature of the underlying process. 
However, this line of work does not permit the prices of traded assets themselves to exhibit fractional dynamics.

In contrast to the existing literature, our stock price process generated by time-changed fBm preserves the principal statistical properties of standard fBm, including its ability to generate long-range dependence, rough sample paths, and anomalous diffusion across parameter regimes. Moreover, its Gaussian--Gamma mixture representation yields a tractable characterization of the process that facilitates both theoretical analysis and statistical implementation. At the same time, it remains fully compatible with the classical semimartingale framework for continuous-time arbitrage-free asset pricing through the use of stochastic activity time, a well-established concept in the option pricing literature.

\section{Stochastic Processes}
\label{sec: Stochastic Processes}

This section introduces the three stochastic processes underlying the fractional Variance Gamma (fVG) model: market point processes, the gamma process, and fractional Brownian motion (fBm). It also establishes several technical results that are used throughout the subsequent development of the model. Throughout the paper, we work on a filtered probability space $(\Omega,\mathcal{F},\mathbb{F}\coloneqq(\mathcal{F}_{t})_{t\in[0,T]},\mathbb{P})$ satisfying the usual hypotheses, where  $\mathbb{P}$ denotes the statistical probability measure. For notational convenience, we suppress explicit dependence on $\omega\in\Omega$ and model parameters whenever no ambiguity arises.

\subsection{Marked point process}
\label{ssec: Stochastic Processes, Marked point process}

Since all pure jump processes can be represented as marked point processes, we provide a brief heuristicoverview of these processes
for readers less familiar with them. For a more comprehensive treatment, we refer readers to standard textbooks, see \citep[e.g.,][]{DaleyVereJones2008TBPointProcess, ContTankov2003TBFinancialModellingJump}.

Pure jump processes are fully characterized by their jump structure. In particular, each jump is identified by its occurrence time and corresponding jump size. While for finite-activity jump processes it is feasible to represent the dynamics explicitly through a sequence of jump-time and jump-size pairs, $\{(T_{1},K_{1}),(T_{2},K_{2}),...\}$, this representation becomes inadequate in the case of infinite-activity processes, where infinitely many jumps may occur over any compact time interval. 

To accommodate this general setting, a more flexible measure-theoretic formulation is required, in which jump times and magnitudes are jointly encoded through random measures that track the full jump structure of the process.
Consider a marked point process $Y$ with associated jump measure
\begin{align*}
	\mu_Y (\diff t, \diff x)  = \sum_{s: \, \Delta Y(s) \neq 0} \delta_{(s,\Delta Y(s))}(\diff t, \diff x) , 
\end{align*} 
where $\delta$ denotes the Dirac measure on $\mathbb{R}_{+}\times\mathbb{R}$
and $\Delta Y(s)$ represents the jump size (mark) of the process $Y$ at time $s$. 
Intuitively, the random measure $\mu_Y$ encodes the jump times and jump size pair of the process $Y$. 
Conversely, given the jump measure $\mu_Y$, the process $Y$ can be recovered via integration against the identity mark function, 
\begin{align*}
	Y(t) = \int_0^t \int_\mathbb{R} x \, \mu_Y(\diff u, \diff x).
\end{align*}

However, the jump measure $\mu_Y$ only encodes the realized jump occurrences and does not directly capture the probabilistic structure of the process. To characterize its distributional properties, it is more convenient to work with its compensator.
$\nu_{Y}$ is defined to be the compensator (measure) of $\mu_{Y}$ if there exists a unique predictable random measure $\nu_Y$ such that, for every Borel set $B \subseteq \mathbb{R}$, the process
\begin{align*}
	t \mapsto \mu_Y((0,t] \times B) - \nu_Y((0,t] \times B)
\end{align*}
is a local martingale, then $\nu_Y$ the compensator of $\mu_Y$. 
It follows that for any nonnegative predictable measurable function $g: [0,T] \times \mathbb{R} \to \mathbb{R}_+$, 
\begin{align*}
	\E\left[\int_s^t \int_\mathbb{R} g(u,x) \, \mu_Y(\diff u, \diff x) \,\middle|\, \mathcal{F}(s) \right] 
	= \E\left[ \int_s^t \int_\mathbb{R} g(u,x) \, \nu_Y(\diff u, \diff x) \,\middle|\, \mathcal{F}(s)\right]
\end{align*}
provided the expectations are well-defined. 
In essence, the compensator $\nu_Y$ governs the predictable (instantaneous) intensity of the jump structure encoded by $\mu_Y$. Analogous to the representation of $Y$ via its jump measure, one can define the compensator-driven finite variation process
\begin{align*}
	A(t)  = \int_0^t \int_\mathbb{R} x \, \nu_Y(\diff u, \diff x)
\end{align*}
such that the process $Y-A$ is a local martingale.

The compensator  $\nu_Y$ characterizes the distributional structure of the process $Y$. 
Informally, for Borel sets $A \subset (0,T]$ and $B \subset \mathbb{R}$, $\nu_Y(A \times B)$ represents the expected number of jumps occurring in the time interval $A$ with jump sizes in $B$. 
If $\nu_Y([0,t] \times \mathbb{R}) = \infty$, the process $Y$ is said to exhibit infinite activity on $[0,t]$; otherwise, it is of finite activity. Analogous to classical analysis, if there exists a measurable function $\psi_Y: [0,T] \times \mathbb{R} \rightarrow \mathbb{R}_+$ such that $\nu_Y(\diff t, \diff x) = \psi_Y(t,x) \diff x \diff t$, then $\nu_Y$ is said to be absolutely continuous with respect to Lebesgue measure, with density/intensity function $\psi_Y$.
In most applied settings involving parametric stochastic processes, the compensator $\nu_Y$ is specified through a parameterized intensity function, which fully determines the statistical behavior of the jump structure.

It is instructive to illustrate the marked point process framework using the canonical example of a Poisson process with constant intensity, even though the general Lévy or jump-diffusion formulations already provide a sufficient theoretical foundation.

\begin{example}[Poisson process: Compensator]
	Consider the Poisson process $N(t)$ with constant intensity $\beta > 0$ (``$\lambda$'' is reserved for its later use as the market price of jump risk).  
	The jump measure of $N$ is
	\begin{align*}
		\mu_N (\diff t, \diff x)  = \sum_{s: \, \Delta Y(s)  = 1} \delta_{(s,1)}(\diff t, \diff x) , 
	\end{align*} 
	The compensator of $\mu_N$ is
	\begin{align*}
		\nu_N(\diff t, \diff x) = \beta \, \delta_1(\diff  x) \diff t ,
	\end{align*}
	where $\delta_1$ is the Dirac delta function centered at $1$.
	Accordingly, the compensator process $A_N$ is
	\begin{align*}
		A_N(t) 
		= \int_0^t \int_\mathbb{R} x \, \nu_N(\diff u, \diff x)
		= \beta \int_0^t \left(\int_\mathbb{R} x \, \delta_1(\diff  x) \right) \diff u
		= \beta t .
	\end{align*}
	and the compensated process $N-A_N = (N(t) - \beta t)_{t \in [0,T]}$ is a martingale.  
	\label{example: Poisson process: Compensator}
\end{example}

We next recall a Girsanov theorem for pure jump processes, following Theorem 3.24 of \citet{JacodShiryaev2003LimitTheoremsStochastic} (Chapter III, p. 172). 
To accommodate the marked point process representation and purely discontinuous dynamics of the time-changed fBm framework considered in this paper, we specialize the result to jump measures by suppressing any continuous martingale component. 

\begin{lemma}[Girsanov theorem for pure jump processes, Theorem 3.24 of \citet{JacodShiryaev2003LimitTheoremsStochastic}]
	Let $Y$ be a local semimartingale, pure jump process with jump measure $\mu_Y$ and compensator $\nu_Y$ under the probability measure $\mathbb{P}$. 
	Let $\lambda: \mathbb{R}_+ \times \mathbb{R} \rightarrow \mathbb{R}_+$ be a predictable process satisfying 
	\begin{align*}
		\lambda(t,x) > 0, \qquad \mathbb{P} \ a.s. \qquad \forall \, t \in [0,T], \qquad x \in \mathbb{R},
	\end{align*}
	and the Novikov's integrability condition
	\begin{align*}
		\E^\mathbb{P} \left[ \exp\left( \frac{1}{2} \int_0^T \int_\mathbb{R} (\lambda(s,x) - 1)^2 \, \nu_Y(\diff s, \diff x)  \right) \right] < \infty .
	\end{align*} 
	Define the probability measure $\mathbb{Q} \ll \mathbb{P}$ by the Radon--Nicodym derivative process
	\begin{align}
		\frac{d \mathbb{Q}}{d \mathbb{P}} \bigg|_{\mathcal{F}(t)} 
		&= \exp\left\{ -\int_0^t \int_\mathbb{R} \left(\lambda(s,x)-1\right) \, \nu_Y(\diff s, \diff x) 
		+ \int_0^t \int_\mathbb{R} \log \lambda(s,x) \, \mu_Y(\diff s,\diff x)   \right\}. 
		\label{eqn: Girsanov theorem for pure jump processes, Radon Nicodym derivative}
	\end{align}
	Then, under $\mathbb{Q}$, the compensator of $\mu_Y$ is given by  
	\begin{align*}
		\tilde{\nu}_Y(\diff t,\diff x) = \lambda(t,x) \, \nu_Y(\diff t,\diff x)
	\end{align*} 
	for all $t \in [0,T]$ and $x \in \mathbb{R}$, so that $Y - \int_0^t \int_\mathbb{R} x \, \tilde{\nu}_Y(\diff s, \diff x) $ is a $\mathbb{Q}$-local-martingale. 
	\label{lemma: Girsanov theorem for pure jump processes}
\end{lemma}

In line with standard terminology, the predictable kernel $\lambda$ is interpreted as the market price of jump risk.
Continuing from Example~\ref{example: Poisson process: Compensator}, we illustrate the application of Girsanov's theorem in Lemma~\ref{lemma: Girsanov theorem for pure jump processes} to the constant-intensity Poisson process.

\begin{example}[Poisson process: Girsanov theorem]
	Define the market price of risk
	\begin{align*}
		\lambda(t,x) &= \frac{\tilde{\beta}}{\beta}
	\end{align*}
	for all $t \in [0,T]$ and $x \in \mathbb{R}$, where $\tilde{\beta} > 0$ is some constant. 
	Then, the Radon Nicodym derivative is 
	\begin{align*}  
		\frac{\diff \mathbb{Q}}{\diff \mathbb{P}} \bigg|_{\mathcal{F}(t)} 
		&=  \exp\left\{ -\int_0^t \int_\mathbb{R} \left(\frac{\tilde{\beta}}{\beta}-1\right) \nu_N(\diff s,\diff x) 
		+ \int_0^t \int_\mathbb{R} \log\left(\frac{\tilde{\beta}}{\beta} \right) \mu_N(\diff s,\diff x)  \ \right\} \\
		&=  \exp\left\{ -  \left(\tilde{\beta}-\beta\right) \int_0^t  \diff s 
		+ N(t) \, \log\left(\frac{\tilde{\beta}}{\beta} \right)   \, \right\} \\
		&= \left(\frac{\tilde{\beta}}{\beta}\right)^{N(t)} e^{-t(\tilde{\beta} - \beta )} 
	\end{align*}
	Under the probability measure $\mathbb{Q}$, $N$ remains a Poisson process with compensator 
	\begin{align*}
		\tilde{\nu}_N = \tilde{\beta} \, \delta_1(\diff x) \diff t  
	\end{align*}
	and the compensator process $\tilde{A}_N$ is 
	\begin{align*}  
		\tilde{A}_N(t) 
		= \int_0^t \int_\mathbb{R} x \, \tilde{\nu}_N (\diff s, \diff x) 
		= \int_0^t \tilde{\beta} \diff s = \tilde{\beta} t .
	\end{align*} 
	Similarly, the compensated process $N - \tilde{A}_N$ is a $\mathbb{Q}$-martingale.
\end{example}


\subsection[Gamma process]{Gamma process, $\gamma$}

Let $\gamma = (\gamma(t; v))_{t \in [0,T]}$ denote a gamma process with unit mean rate and variance intensity $v$, characterized by independent Gamma-distributed increments over non-overlapping intervals \citep{MadanCarrChang1998VGoption}.\footnote{
	It is straightforward to generalize the specification by introducing an additional parameter that allows the expected activity rate to differ from calendar time. However, such a generalization lacks a clear economic motivation. Accordingly, we normalize the mean activity rate to unity.
}
The gamma process will subsequently serve as a stochastic time change, referred to as activity time, to distinguish it from calendar time.
For any time $t\geq 0$ and lag $h \geq 0$, the increment $G_h := \gamma(t+h; v) - \gamma(t; v)$ follows a Gamma distribution with shape parameter $h/v$ and rate parameter $1/v$, or equivalently, the probability density function of $G_h$ is given by
\begin{align}
	f_{G_h}(g) = \frac{g^{h/v-1} \, e^{-g/v}}{\Gamma(h/v) \, v^{h/v}} ,
	\qquad \forall \, g > 0, \, h \geq 0
	\label{eqn: Background, pdf of G_h}
\end{align} 
where $\Gamma(x) = \int_0^\infty z^{x-1} e^{-z} \diff z$ denotes the gamma function for all $x \in \mathbb{R} \backslash \{0,-1,-2,...\}$. 
The increments have expectation $\E[G_h] = h$ and variance $\Var[G_h] = vh$. 
Intuitively, when viewed as a stochastic time change (or subordinator), the gamma activity time evolves at unit speed in expectation relative to calendar time. The volatility parameter $v$ governs the variability of the activity clock around its mean and therefore controls the randomness in the accumulation of activity time per unit of calendar time.

The gamma process $\gamma$ is a non-decreasing Lévy process with independent and stationary increments, and is therefore referred to as a subordinator. Moreover, it is an infinite-activity (infinitely many jumps on any finite interval), pure-jump (purely discontinuous) process.
The gamma process admits the following marked point process representation\footnote{
	While the Lévy measure is often sufficient in the Lévy process setting, we adopt the marked point process formulation because it extends naturally to the subsequent analysis. In particular, the time-changed fBm introduced later is not a Lévy process due to the loss of independent increments. The compensator-based representation therefore provides the appropriate framework for characterizing its jump structure.}
with absolutely continuous compensator $\nu_{\gamma}$ satisfying
\begin{align*}
	\nu_{\gamma}(\diff t, \diff g)  = \psi_{\gamma}(g) \diff g  \diff t ,
	\qquad \forall \, g > 0, t \in [0,T],
\end{align*}
where the compensator density function is given by
\begin{align}
	\psi_{\gamma}(g) = \frac{e^{-g/v}}{v g} ,
	\qquad \forall \, g >0 .
	\label{eqn: Background, compensator density of gamma activity}
\end{align}

\subsection[The fBm process]{The fBm process, $B_{H}$}

The standard (one-sided) fBm $B_{H}= (B_H(t))_{t\in[0,T]}$ is a centered continuous Gaussian process with stationary increments, initial condition $B_H(0) = 0$, and covariance function
\begin{align*}
	\E[B_H(t) B_H(s)] = \frac{1}{2} \left(t^{2H} + s^{2H} - |t-s|^{2H} \right), 
	\qquad \forall \, s,t \in [0,T],
\end{align*}
where $H \in (0,1)$ denotes the Hurst coefficient. 
The process is $H$-self-similar, satisfying $B_H(ct) \stackrel{d}{=} |c|^H B_H(t)$ for all $c \in \mathbb{R}$, and possesses sample paths that are almost surely $\alpha$-Hölder continuous for any order $\alpha \in [0,H)$.  When $H = 0.5$, the fBm reverts to the classical Brownian motion.

Distinct from classical Brownian motion, the fBm exhibits dependent increments. When $H>0.5$, $B_{H}$ admits smoother sample paths than standard Brownian motion, and its increments possess positive serial correlation and display long-range dependence (LRD), characterized by the hyperbolic rate of decay in autocorrelations such that distant observations retain non-negligible influence on future dynamics.\footnote{
	Several notions of LRD are used in the literature. A standard definition characterizes a covariance-stationary process $\{X_k\}_{k \in \mathbb{Z}}$ as exhibiting LRD if its autocovariance function $\{\gamma_X(k)\}_{k \in \mathbb{Z}}$ is not absolutely summable, i.e., $\sum_{k=-\infty}^\infty |\gamma_X(k)| = \infty$.
	Alternatively, LRD can be characterized by the asymptotic rate of  decay of the autocorrelation function, satisfying $\gamma_X(k) \sim k^{2H-2}$ as $k \to \infty$, or, in the frequency domain, via the divergence of the spectral density at the origin, with the spectral density behaving as $\lambda^{-(2H-1)}$ as $\lambda \to 0$ for $H \in (0.5, 1)$.
	The increments of the fBm process satisfy all these definitions when $H > 0.5$.}  
This contrasts sharply with classical Brownian motion, whose increments are independent and Markovian, and thus exhibit no temporal persistence. Such dependence structures are particularly relevant for modeling persistent  serial correlation.
In contrast, when $H<0.5$, the increments are negatively autocorrelated, and the process exhibits comparatively rough sample paths. 
Crucially, when $H \neq 0.5$, the fBm process ceases to be a semimartingale, which precludes the direct application of standard Itô calculus and risk-neutral valuation framework.

Furthermore, the fBm exhibits anomalous diffusion, as whereby its variance scales nonlinearly over time, $\Var[B_H(t)] \propto t^{2H}$, deviating from the linear time scaling of standard Brownian motion. Accordingly, the process is superdiffusive for $H > 0.5$, with variance growing faster than linearly in time, and subdiffusive for $H < 0.5$, with slower-than-linear growth. In addition, the Hurst parameter governs path regularity; smaller values of $H$ correspond to rougher trajectories with weaker Hölder continuity, while larger values of $H$ yield progressively smoother sample paths.

A particularly useful characterization of fractional Brownian motion is its Volterra integral representation in terms of an underlying Brownian motion. Specifically,
\begin{align}
	B_H(t) = \int^t_0 K_H(t,u) \diff B(u), 
	\qquad \forall \, t \in [0,T],
	\label{eqn: Background, Volterra representation of fBm}
\end{align}
where $B = (B(t))_{t \in [0,T]}$ is a standard Brownian motion and $K_H$ is a deterministic kernel given by 
\begin{align}
	K_H(t,u) 
	&= \sigma_H \frac{(t-u)^{H-0.5}}{\Gamma\left(H + \frac{1}{2}\right)} \left[ \left(H-\frac{1}{2}\right) \int_0^1 v^{H-\frac{3}{2}} \left(1-\left(1-\frac{t}{u}\right)v\right)^{H-\frac{1}{2}} \diff v\right] 1_{[0,t)}(u)
	\label{eqn: Background, K_H} \\ 
	\sigma_H^2 
	&= \frac{2 \pi H\left(H - \frac{1}{2}\right)}{\Gamma(2-2H) \sin\left(\pi \left(H - \frac{1}{2}\right)\right)} 
	\label{eqn: Background, sigma_H}   
\end{align} 
for all $t,u \in [0,T]$. 
The filtration generated by $B_H$ coincides with that generated by the driving Brownian motion $B$, $\sigma\left((B_H(u))_{u \in [0,t]}\right)=\sigma\left((B(u))_{u \in [0,t]}\right)$ for all $t \in [0,T]$.   
For further technical details on fBm, we refer the readers to \citet{MandelbrotVanNess1968fBm, SamoradnitskyTaqqu1994StableProcess, DecreusefondUstunel1998fBm}.

In parallel with the classical Brownian motion, a version of Girsanov's theorem holds for fBm.
\begin{lemma}[Girsanov Theorem for fBm, \citet{DecreusefondUstunel1998fBm}]
	Let $(\lambda(t))_{t\in[0,T]}$ be a square-integrable, adapted process
	with respect to Lebesgue measure such that $\E^{\mathbb{P}}[\mathcal{L}(T)]=1$,
	where 
	\begin{align*}
		\mathcal{L}(t)=\exp\left\{ -\int_{0}^{t}\lambda(v)\diff B(v)-\frac{1}{2}\int_{0}^{t}\lambda(v)^{2}\diff v\right\}, 
		\qquad \forall \, t\in[0,T].
	\end{align*} 
	Define the equivalent probability measure $\mathbb{Q}$ by the Radon-Nikodym
	derivative $\frac{\diff\mathbb{Q}}{\diff\mathbb{P}}|_{\mathcal{F}(t)}=L(t)$. 
	Then, $(B^{H}(t)+\int_{0}^{t}K_{H}(t,u)\lambda(u)\diff u)_{t\in[0,T]}$
	under $\mathbb{P}$ has the same law as $(B^{H}(t))_{t\in[0,T]}$
	under $\mathbb{Q}$. 
	\label{lemma: fractional Girsanov theorem by DecreusefondUstunel1998fBm}
\end{lemma}

Owing to its analytical tractability and ability to capture both persistence and roughness, fBm has been widely adopted across disciplines, including hydrology \citep{Hurst1951Reservior, MandelbrotWallis1968Joseph}, 
climate science \citep{Koutsoyiannis2003ClimateChgHurst, Iliopoulou2018LRDPrecipitation, PelletierTurcotte1997LRDHydrologyClimatology, Franzke2010LRDAntarctic, Smith1993LRD},
operations \citep{AbryVeitch1998LRDTraffic, GrossglauserBolot1999LRDTraffic, Park2011LRDInternetTraffic, LelandTaqquWillingerWilson1993LRDEthernet}, and 
life sciences \citep{Tagliazucchi2013LRDSleep, Heath1998LRDFluid, PengEtAl1992LRDinDNA}. 
In finance, fBm has been extensively studied in relation to long memory in asset returns \citep{GreeneFielitz1977LRDStockReturns, Willinger1999LRDFinance, HuOksendal2003fractionalWhiteNoise} and the roughness and clustering of stochastic volatility \citep{Wang2023volatility, RayTsay2000LRDVolatility}. In actuarial science and insurance, it has been used to model temperature and mortality dynamics for pricing weather derivatives and life-contingent claims \citep{Tan2026PersistentTempContracts, WangEtAl2021VolterraMortalityModel, Zhou2022MortatlitymfBm}.

We now introduce notation that will be used in the construction of the time-changed fBm in the following section. Let $\phi(\cdot \, ; t,g)$ denote the conditional probability density function (PDF) of the increment $B_H(t+g) - B_H(t)$ given $\mathcal{F}(t)$, for $t \in [0,T]$ and lag $g \geq 0$,
\begin{align}
	\phi(x;t,g) 
	& =\frac{1}{\sqrt{2\pi}\sigma_{B}(g;t)}\exp\left\{ -\frac{1}{2}\left(\frac{x-m_{B}(g;t)}{\sigma_{B}(g;t)}\right)^{2}\right\} ,
	\qquad \forall \, x \in \mathbb{R},
	\label{eqn: Background, pdf of conditional fBm}
\end{align}
where $m_{B}(g; t)$ and $\sigma_{B}^{2}(g;t)$ are the conditional mean and variance of the increment $B_H(t+g)-B_H(t)$,
\begin{align}
	m_{B}(g;t) 
	&= \E\left[B_H(t+g)-B_H(t)|\mathcal{F}(t)\right]
	=\int_{0}^{t}K_{H}(t+g,u) \, \diff B(u) 
	\label{eqn: Background, conditional mean of fBm increment mB} \\
	\sigma_{B}^{2}(g;t) 
	& = \Var\left[B_H(t+g)-B_H(t)|\mathcal{F}(t)\right]
	=\int_{t}^{t+g}K_{H}(t+g,u)^{2} \, \diff u .
	\label{eqn: Background, conditional variance of fBm increment sigma2B} 
\end{align}

\section[Time-changed fBm process]{Time-changed fBm process, $X=B_{H}(\gamma)$}
\label{sec: Time-changed fBm process}

This section introduces the time-changed fBm $X$ underlying the fVG model. The process $X = (X(t))_{t \in [0,T]}$ is obtained by evaluating fBm $B_H$ at a gamma time change $\gamma$, where $B_H$ and $\gamma$ are assumed to be independent. Specifically,
\begin{align*}
	X(t) \equiv X(t; H, v) = B_H(\gamma(t; v)),
	\qquad \forall \, t \in [0,T], \, v \geq 0, \, H \in (0,1).
\end{align*}

\subsection{Properties} 
\label{ssec: time changed fBm, Properties}

We begin by summarizing several key properties of the time-changed fBm process established in the literature
\citep{Kozubowski2006fLaplacemotion, Kumar2017TimeChgfBm}.
First, the probability density function (PDF) of the time-changed fBm process $X$ is derived by marginalizing the Gaussian conditional PDF of the fBm over the Gamma-distributed PDF governing the increments of the activity time process, 
\begin{align}
	f_{X(t)}(x)
	&= \int_0^\infty \phi(x; 0, g) \, f_{G_t}(g) \diff g \nonumber \\
	&=\frac{1}{\sqrt{2\pi} \, v^{t/v}\, \Gamma\left(t/v\right)}\int_{0}^{\infty} g^{t/v-H-1} e^{- x^{2} g^{-2H}/2-g/v} \diff g,\qquad \forall \, x \in \mathbb{R}\backslash \{0\},
	\label{eqn: Time changed fBm process, unconditional pdf of X}
\end{align}
where $f_{G_t}$ and $\phi$ are defined in Equations~\eqref{eqn: Background, pdf of G_h} and~\eqref{eqn: Background, pdf of conditional fBm}, respectively.
Second, the covariance function of $X$ takes the form 
\begin{align*}
	\E\left[X(s) X(t)\right]
	=\frac{v^{2H}}{2}\left( \frac{\Gamma(\frac{t}{v}+2H)}{\Gamma(\frac{t}{v})}+\frac{\Gamma(\frac{s}{v}+2H)}{\Gamma(\frac{s}{v})}-\frac{\Gamma(\frac{|t-s|}{v}+2H)}{\Gamma(\frac{|t-s|}{v})}\right),
	\qquad \forall \, s,t \in [0,T].
\end{align*}
Importantly, the gamma activity time change preserves the long-memory properties of the underlying fBm.
The covariance function of the non-overlapping increments of $X$ is 
\begin{align} 
	&\E\left[\left( X(t + (n+1)h) - X(t+nh) \right) (X(t+h) - X(t) ) \right] \nonumber \\
	&\qquad = \frac{v^{2H}}{2} \left(
		\frac{\Gamma(\frac{(n-1)h}{v}+2H)}{\Gamma(\frac{(n-1)h}{v})}
		- 2 \frac{\Gamma(\frac{nh}{v}+2H)}{\Gamma(\frac{nh}{v})}
		+ \frac{\Gamma(\frac{(n+1)h}{v}+2H)}{\Gamma(\frac{(n+1)h}{v})}
	\right)  ,
	\quad \forall \, n \in \mathbb{N}, \, h > 0,
	\label{eqn: time changed fBm, autocovariance of increments of X}
\end{align}
and admits the asymptotic behavior $\rho_h(n) \sim H(2H-1) h^{2H} v^{-2H} n^{2H-2}$ as $n \rightarrow \infty$.
Consequently, the increment process exhibits long-range dependence (LRD) whenever $H > 0.5$.

\begin{figure}[!htp]
	\centering
	\includegraphics[width = .95\linewidth]{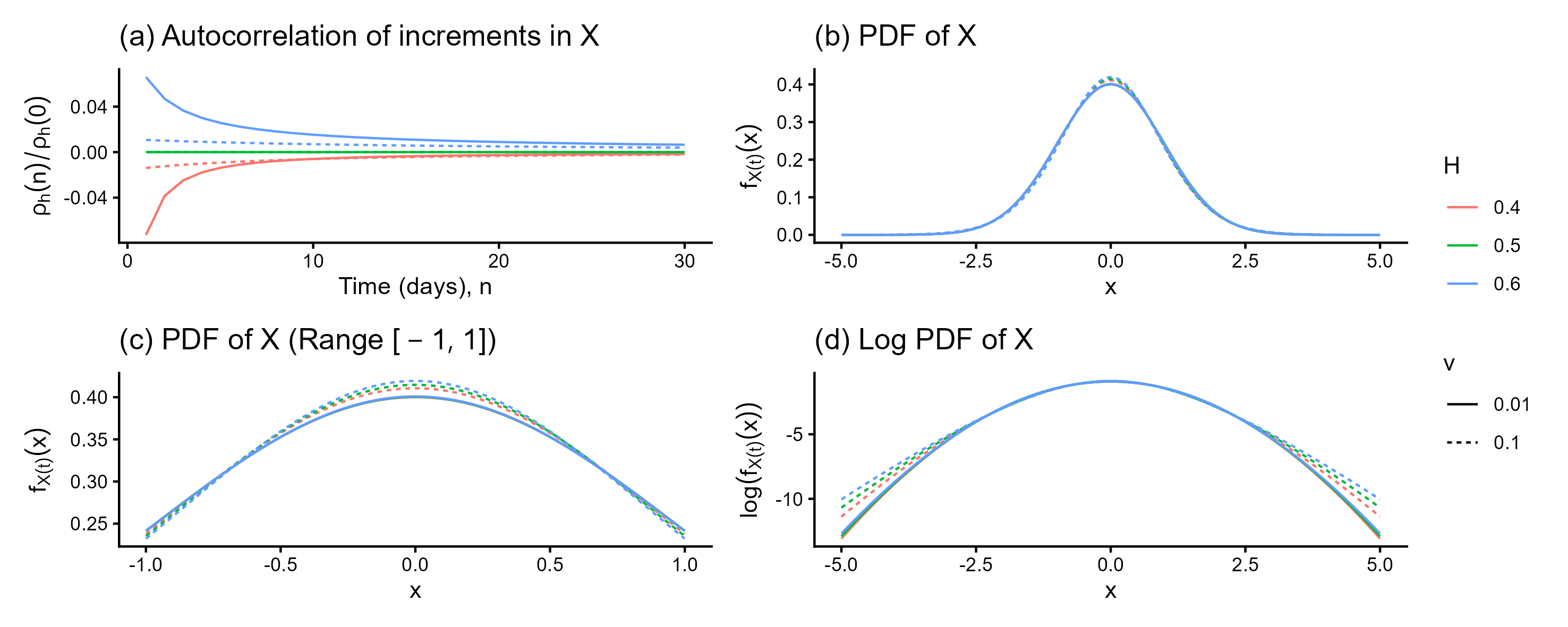}
	\caption{Autocorrelation of the increments in $X$ and PDF of $X$ across various $v,H$}
	\label{fig: time changed fBm, Autocorrelation and PDF of time changed fBm}
\end{figure}

Figure~\ref{fig: time changed fBm, Autocorrelation and PDF of time changed fBm}(a) plots the autocorrelation function $\rho_h(n) / \rho_h(0)$ for various combinations of $H$ and $v$ up to lag $n=30$, where $h=1/252$ corresponds to the average length of a trading day expressed in years and $\rho_h(n)$ is given by Equation~\eqref{eqn: time changed fBm, autocovariance of increments of X}. 
The autocorrelation profiles indicate the presence of LRD in the increments of $X$ when $H > 0.5$, as demonstrated by the slow decay of the autocorrelation function. In addition, increasing the variance intensity parameter $v$ reduces the magnitude of the autocorrelations, aside when $n$ is large and $H < 0.5$.

Figure~\ref{fig: time changed fBm, Autocorrelation and PDF of time changed fBm}(b) shows the PDF of $X(1)$, where $f_{X(t)}(x)$ is given in Equation~\eqref{eqn: Time changed fBm process, unconditional pdf of X}. 
Since the overall scale may obscure important features of the distribution, two supplementary visualizations are provided. Figure~\ref{fig: time changed fBm, Autocorrelation and PDF of time changed fBm}(c) displays the same PDF with the domain restricted to $x \in [-1,1]$, thereby emphasizing the behavior around the peak, while Figure~\ref{fig: time changed fBm, Autocorrelation and PDF of time changed fBm}(d) presents the logarithm of the PDF to better illustrate the tail behavior.
The results show that larger values of $H$ and $v$ lead to a lower peak and heavier tails, indicating an increase in dispersion and tail thickness. However, when $v$ is small, variations in $H$ have only a limited effect on the shape of the PDF. This suggests that $v$ is the primary parameter governing the kurtosis of the time-changed fBm process.

Next, we derive the unconditional moments of the time-changed fBm process in Proposition \ref{prop: time changed fBm, unconditional raw moments}. 

\begin{prop}[Unconditional raw moments of $X$]
	For any $n \in \mathbb{N}$, 
	\begin{align}
		\E[B^H(\gamma(t))^n ] =  \begin{cases}
			v^{nH} \frac{\Gamma(t/v + nH)}{\Gamma(t/v)}  (n-1)!! , & n \text{ even} \\
			0, & n \text{ odd}
		\end{cases}.
	\end{align}
	\label{prop: time changed fBm, unconditional raw moments}
\end{prop}

It follows from Proposition~\ref{prop: time changed fBm, unconditional raw moments} that the unconditional mean and skewness of $X$ are both zero.
Moreover, the unconditional variance is given by $\Var[X(t)] = v^{2H} \frac{\Gamma(t/v + 2H)}{\Gamma(t/v)} $, and the unconditional kurtosis is 
\begin{align*}
	\frac{\E[(X(t) - \E[X(t)])^4]}{\Var[X(t)]^2}
	&= \frac{3 v^{4H} \Gamma(t/v + 4H)/\Gamma(t/v)}{(v^{2H} \Gamma(t/v + 2H)/\Gamma(t/v))^2}
	= 3 \frac{\Gamma(t/v + 4H)\Gamma(t/v)}{\Gamma(t/v + 2H)^2} \\
	&= \frac{3}{\prescript{}{1}{F}_2(-2H,2H;\frac{h}{v}+2H; 1)}
	= 3 \prod_{k=0}^\infty \frac{\left(\frac{h}{v} + 2H + k \right)^2}{\left(\frac{h}{v} + 4H + k \right) \left(\frac{h}{v} + k \right)},
	\qquad \forall \, t \geq 0, \, v > 0.
\end{align*}
Thus, $X$ is leptokurtic, and the variance intensity parameter $v$ controls the kurtosis of $X$. As $ t \rightarrow \infty$, the kurtosis decreases and converges to 3, corresponding to the kurtosis of a Gaussian process.
These results are consistent with the observations shown in Figure~\ref{fig: time changed fBm, Autocorrelation and PDF of time changed fBm} and with the findings of  \citet{Kozubowski2006fLaplacemotion} and, in the special case $H=0.5$, of \citet{MadanCarrChang1998VGoption}.

Next, we describe the simulation procedure for the time-changed fBm $X$ on the interval $[0,T]$ with time step $a > 0$, given its parameters $v\geq0$ and $H\in(0,1)$. The idea is to first simulate both the discretized fBm and the gamma processes independently, and then evaluate the fBm at the corresponding realized activity times generated by the gamma process.
To ensure numerical accuracy, the fBm is generated on a finer (secondary) grid than that used for the gamma activity time (primary grid).

Let the step sizes of the primary and secondary grids be $a$ and $b$, respectively, with $a > b > 0$. 
The primary grid $\{t_n\}_{n=1}^N$ is constructed by $t_n = n a$, where $N = \lfloor T/a \rfloor$ denotes the number of primary grid points. 
We first simulate the gamma process $\{\hat{\gamma}_n\}_{n=1}^N$ with $\hat{\gamma}_n = \gamma(t_n)$. 
Subsequently, we construct the secondary grid $\{s_j\}_{j=1}^{J}$ via $s_j = jb$, where $J = \lfloor \hat{\gamma}_N/b \rfloor$ represents the number of secondary grid points.
Next, we generate a sample path of the fBm $\{\hat{B}_{H,j}\}_{j=1}^J$ on the secondary grid, with $\hat{B}_{H,k} = B_H(s_j)$.
We then map the realized activity times onto the fBm path by selecting $\hat{B}_{H,n} = \hat{B}_{H,j}$ with $j = \lfloor \hat{\gamma}_n / b \rfloor$ for each $n \in \{1,...,N\}$. 
The resulting sequence  $\{\hat{B}_{H,n}\}_{n=1}^N$ constitutes a single simulated path of the time-changed fBm process $X$. 
Throughout the paper, we set $a = 1/252$, corresponding to the length of a trading day measured in years, and $b= a/100$.

For simulation of the gamma process, we draw independent and identically distributed gamma increments with shape parameter $a/v$ and rate parameter $1/v$, and construct the process via cumulative summation. The fBm paths are generated by summing simulated increments obtained through the Wood–Chan method, which exploits a circulant embedding of the covariance matrix of fractional Gaussian noise and employs the Fast Fourier Transform for computational efficiency \citep{WoodChan1994SimulateGaussian}.

\begin{figure}[!htp]
	\centering
	\includegraphics[width = .95\linewidth]{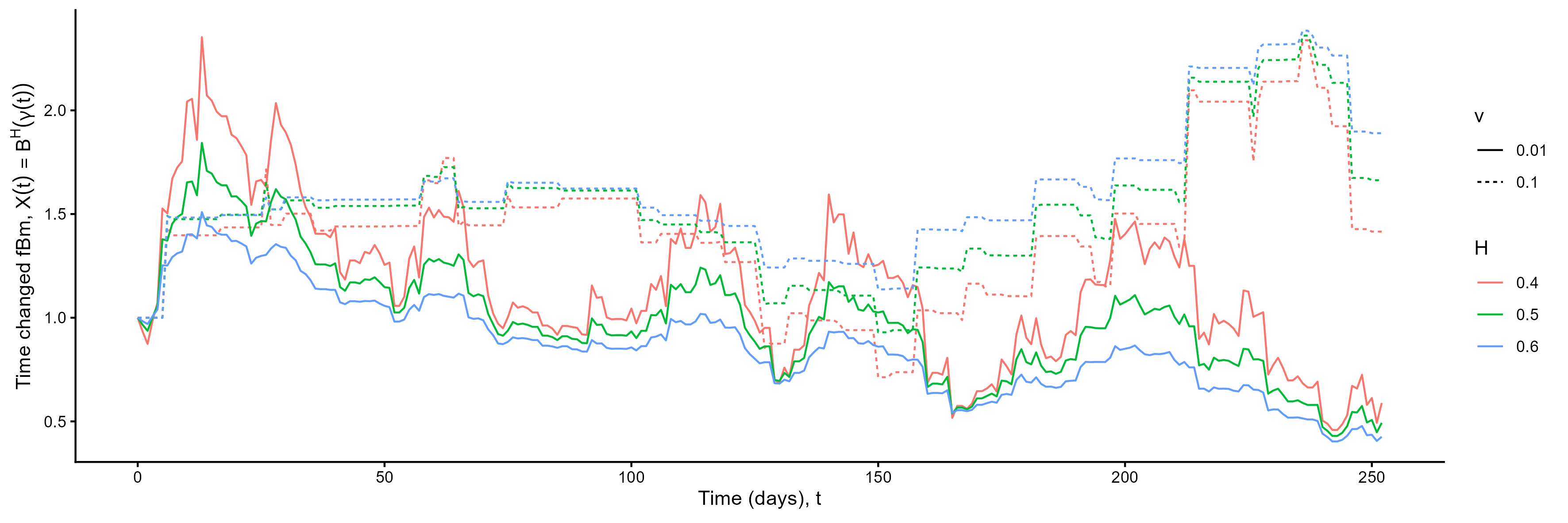}
	\caption{Simulated sample paths of $X$ under various $v,H$}
	\label{fig: time changed fBm, Simulated sample paths}
\end{figure}

Figure~\ref{fig: time changed fBm, Simulated sample paths} presents simulated sample paths of time-changed fBm on the interval $[0,1]$ under various combinations of the parameters $\{v,H\}$. To facilitate direct comparison across parameter settings, the same random seed is used in all simulations, thereby eliminating sampling variability and isolating the effects of $v$ and $H$ on the sample paths.
When $H < 0.5$, the sample paths exhibit a relatively rough and irregular appearance, whereas for $H > 0.5$, the trajectories become noticeably smoother. This reflects the fact that the Hurst exponent governs the local regularity of the sample paths, a property inherited from the underlying fBm.

In contrast, as $v$ increases, the trajectories appear increasingly intermittent, characterized by extended periods of little or no movement followed by abrupt changes. This phenomenon arises because a larger variance intensity parameter induces greater variability in the increments of the gamma activity time. Although the mean rate of the gamma activity time remains unchanged, its increment distribution becomes more positively skewed and heavy-tailed, resulting in more frequent occurrences of both very small and exceptionally large increments. Consequently, the time-changed process exhibits prolonged flat segments interspersed with sudden, pronounced movements.
Overall, the results indicate that the Hurst exponent $H$ retains its classical interpretation as a measure of path roughness in the time-changed fBm model, while the variance intensity parameter $v$ primarily influences the intermittency and irregular timing of the process evolution.

\subsection{Marked point representation}

To characterize the stochastic drift of the time-changed fBm process $X$, it is more convenient to represent $X$ as a marked point process. Since fBm has continuous sample paths and the gamma activity time is an infinite activity, pure jump process, the resulting time-changed process $X$ is also a purely discontinuous, pure jump process.

Let $\nu_X$ denote the compensator of $X$ and let $\psi_X$ be the associated density of $\nu_X$, satisfying 
\begin{align}
	\nu_X(\diff t, \diff x) &= \psi_X(t,x) \, \diff x \, \diff t ,
	& \forall \, t \in [0,T], x \in \mathbb{R} .
	\label{eqn: time changed fBm, compensator measure for X}
\end{align}
By the construction of $X$, the compensator density $\psi_X$ can be expressed as the convolution of the conditional density of the fractional Brownian motion increments with the density of the gamma activity process,
\begin{align}
	\psi_X(t,x) 
		&=\int_{0}^{\infty}\phi(x; \gamma(t; v),g) \, \psi_{\gamma}(g) \, dg 
		\nonumber\\
		& =\int_{0}^{\infty}\frac{\exp\left\{ -\frac{1}{2}\left(\frac{x-m_{B}(g;\gamma(t; v))}{\sigma_{B}(g;\gamma(t; v))}\right)^{2}\right\} }{\sqrt{2\pi}\sigma_{B}(g;\gamma(t; v))}\cdot\frac{1}{v g}e^{-\frac{g}{v}} \, dg,
		& \forall \, t \in [0,T], x \in \mathbb{R},
	\label{eqn: time changed fBm, density of jump measure of X}
\end{align}
where the functions  $\psi_{\gamma}$, $\phi$, $m_{B}$ and $\sigma_{B}$ are defined in Equations~\eqref{eqn: Background, compensator density of gamma activity}, \eqref{eqn: Background, pdf of conditional fBm}, \eqref{eqn: Background, conditional mean of fBm increment mB}, and \eqref{eqn: Background, conditional variance of fBm increment sigma2B}, respectively.
The conditional moments $m_{B}$ and $\sigma_{B}$ are evaluated at the latent gamma activity time $\gamma(t)$, 
\begin{align*}
	m_{B}(g;\gamma(t)) 
	& =\E\left[B_{H}(\gamma(t)+g)-B_{H}(\gamma(t))|\mathcal{F}(t-)\right]
	=\int_{0}^{\gamma(t)}K_{H}(\gamma(t)+g,u) \, \diff B(u)\\
	\sigma_{B}^{2}(g;\gamma(t)) 
	& =\Var\left[B_{H}(\gamma(t)+g)-B_{H}(\gamma(t))|\mathcal{F}(t-)\right]
	=\int_{\gamma(t)}^{\gamma(t)+g}K_{H}(\gamma(t)+g,u)^{2} \, \diff u
\end{align*}
for all $g \geq 0$ and $t \in [0,T]$.

\subsection{Semimartingale Behavior and Equivalent Probability Measure Changes}
\label{ssec: Time-changed fBm process, Semimartingale Behavior and Equivalent Probability Measure Changes}

This section presents the main theoretical results establishing the suitability of the time-changed fBm $X$ as a driving process for asset price models within the classical arbitrage-free pricing framework. We first prove that the time-changed fBm is a semimartingale and then characterize its compensator.

\begin{theorem}[Semimartingale $X$]
	The time-changed fBm $X = B_H(\gamma)$ is a semimartingale.
	\label{thm: time changed fBm, semimartingale X}
\end{theorem}

\begin{prop}[Compensator of $X$]
	There exists a unique, continuous, predictable, finite variation process $A_X = \int_0^t a_X(s) \diff s$ such that $X-A_X$ is a $\mathbb{P}$-martingale, where
	\begin{align*}
		a_X(s) = \int_{\mathbb{R}} x \, \psi_{X}(s,x) \, \diff x = \int_{0}^{\infty}m_{B}(g;\gamma(s)) \, \psi_{\gamma}(g) \, \diff g ,
		\qquad \forall \, s \in [0,t]. 
	\end{align*} 
	\label{prop: time changed fBm, Compensator of X}
\end{prop}

Theorem~\ref{thm: time changed fBm, semimartingale X} establishes that the gamma time-changed fBm X is a semimartingale, while Proposition~\ref{prop: time changed fBm, Compensator of X} provides an explicit characterization of its predictable compensator. In particular, the compensator is unique, and the compensated process $X-A_X$ is a true $\mathbb{P}$-martingale. 
Intuitively, the process $a_{X}=(a_{X}(t))_{t\in[0,T]}$ represents the instantaneous predictable drift of the time-changed fBm, in the sense that $a(t)\approx\E[B_{H}(\gamma(t+\diff t))-B_{H}(\gamma(t))|\mathcal{F}(t-)] / \diff t$, which quantifies the conditional expected infinitesimal rate of change given the available information.
Under the special case $H=0.5$, the underlying fBm reduces to a standard Brownian motion, whose increments have zero conditional mean. In	this case, $a_{X}(t)=0$ for all $t\in[0,t]$, and therefore, the compensator process vanishes, $A_{X}(t)=0$, and the time-changed Brownian motion is a martingale.

Consequently, as discussed in the introduction, $X$ may be employed as the driving noise process in asset price models within the classical continuous-time, continuous-trading, arbitrage-free pricing framework, thereby overcoming the principal obstacle to the direct use of fBm, namely its failure to be a semimartingale. Moreover, the existence of a unique predictable compensator facilitates the derivation of sufficient conditions on the drift of an asset price process under which an equivalent change of probability measure renders the discounted price process a martingale. The following corollary establishes Girsanov's change-of-measure theorem for the time-changed fBm.

\begin{coro}[Girsanov theorem for time-changed fBm]
	Consider a time-changed fBm process $X=B_H(\gamma)$ with compensator $\nu_X$ under the statistical measure $\mathbb{P}$.
	Suppose there exists a predictable process $\lambda > 0$ satisfying all the necessary conditions in Lemma \ref{lemma: Girsanov theorem for pure jump processes}, and define the probability measure $\mathbb{Q}$ by the Radon-Nikodym derivative in Equation~\eqref{eqn: Girsanov theorem for pure jump processes, Radon Nicodym derivative}. 
	Then, there exists a unique, predictable, finite-variation compensator process  
	$\tilde{A}_X(t) = \int_0^t \int_\mathbb{R} x \, \lambda(s,x) \, \nu_X(\diff s, \diff x) $
	such that the compensated process $X - \tilde{A}_X$ is a $\mathbb{Q}$-local-martingale. 
	Furthermore, if 
	\begin{align}
		\E^\mathbb{P} \left[ \int_0^T \int_\mathbb{R} \, x^2 \, \lambda(s,x) \, \psi_{X}(s,x) \, \diff x \, \diff s \, \right] < \infty,
		\label{eqn: time changed fBm, compensated X martingale, condition on MPR}
	\end{align}
	then $X - \tilde{A}_X$ is a $\mathbb{Q}$-martingale. 
	\label{coro: Girsanov theorem for time changed fBm}
\end{coro}

Corollary~\ref{coro: Girsanov theorem for time changed fBm} characterizes the predictable compensator of the time-changed fBm under an equivalent risk-neutral measure. A sufficient condition for Equation~\eqref{eqn: time changed fBm, compensated X martingale, condition on MPR} is that the market price of risk be uniformly bounded, that is, $|\lambda(t,x)| < M$ for some constant $M > 0$. Indeed, since the compensator under $\mathbb{Q}$ satisfies $\tilde{\nu}=\lambda\nu$, and the expected quadratic variation of $X$ under $\mathbb{P}$ is finite, it follows that 
\begin{align*}
	\E^\mathbb{P} \left[ \int_0^T \int_\mathbb{R} x^2 \lambda(s,x) \psi_{X}(s,x) \diff x \diff s \right]
	< M \E^\mathbb{P} \left[ \int_0^T \int_\mathbb{R} x^2 \psi_{X}(s,x) \diff x \diff s \right]
	< \infty.
\end{align*}

Finally, we highlight some challenges arising in the application of time-changed fBm within financial modeling. In particular, the conditional distribution of $X(t)$ given $\mathcal{F}(s)$, for $s \leq t$, cannot be determined solely based on the realizations of $X$ up to time $s$. Although its conditional law admits a representation in terms of the latent gamma activity time $\gamma(s)$, the realized values of $\gamma(s)$ are unobservable. Specifically, under the natural filtration $\mathcal{F}(s) = \sigma(\{X(u) \, : \, u \in [0,s]\} )$, the activity time $\gamma(s)$ is not $\mathcal{F}(s)$-measurable.
Consequently,the constituent processes $B_H$ and $\gamma$ cannot be separately identified from the observations of $X=B_H(\gamma)$.
Likewise, the compensator process $A_X$ is not adapted to the filtration generated by $X$, since it depends on the conditional moments $m_{B}(g,\gamma(s))$ and $\sigma_{B}^{2}(g;\gamma(s))$, which are functions of the latent activity time $\gamma$ and therefore cannot be evaluated directly from the observed sample path of $X$.

Although the conditional distribution of $X$ is unavailable in closed form, it can be approximated by simulation. The Gaussian--Gamma mixture representation of the time-changed fBm makes unconditional path simulation of $X$ computationally straightforward. The conditional distribution can then be approximated by retaining only those simulated paths that are sufficiently close to the observed history. 
Specifically, given historical observations $\{\hat{X}(t_k)\}_{k=0}^K$ with $0 = t_0 < t_1 < \dots < t_K = s$ and parameters $\{v,H\}$, the conditional distribution may be approximated by
\begin{align*}
	\hat{F}_{X(t) | \mathcal{F}(s)}(x) = \frac{\sum_{i=1}^N 1\left[X^{(i)}(t) \leq x \,\middle|\, | X^{(i)}(t_k) - \hat{X}(t_k)| < \delta_k \quad \forall \, k\right]}{\sum_{i=1}^N 1\left[| X^{(i)}(t_k) - \hat{X}(t_k)| < \delta_k \quad \forall \, k \right]} 
\end{align*}
which converges to the true conditional distribution as $N \rightarrow \infty$ and $\delta_k \rightarrow 0$ for all $k$. Simulation of $X$ is discussed in Section~\ref{ssec: time changed fBm, Properties}, while parameter estimation is presented in Section~\ref{sec: Parameter Estimation}. Conditional quantities of practical interest, such as forecasts based on conditional expectations and measures of future uncertainty based on conditional variances, can then be evaluated directly from the simulated distribution.

\subsection{Skewness}
\label{ssec: time changed fBm, Skewness}

The standard time-changed fBm $X$ is symmetric and therefore cannot capture
skewness are commonly observed in financial return series. In this section, we introduce an extension analogous to the VG model of \citet{MadanCarrChang1998VGoption} by appending an additional drift term $\theta \gamma(t; v)$ to introduce asymmetry, thereby enabling the model to capture skewed return distributions.

Consider the process $W = (W(t))_{t \in [0,T]}$, where
\begin{align*}
	W(t) \equiv W(t; \theta, \sigma, v, H) = \theta \gamma(t; v) + \sigma X(t; H, v).
\end{align*}
The four parameters in $W$ have distinct distributional effects. The parameter $\theta \in \mathbb{R}$ primarily controls the mean and skewness of the increments, $v \geq 0$ controls the variability of the activity time and hence the excess kurtosis, $\sigma \geq 0$ determines the overall scale of the process, and $H \in (0,1)$ governs the path regularity, degree of persistence, and the rate at which moments scale with the time horizon. This separation of roles is useful for estimation, as changes in skewness, kurtosis, scale, and temporal dependence can be attributed to different parameters of the model.
The proposed specification represents a fractional extension of the classical VG model. When $H=0.5$, the model coincides with the VG process. For $H\neq 0.5$, the model retains the variance-gamma mechanism responsible for skewness and excess kurtosis while additionally incorporating long-range dependence and fractional scaling through the Hurst parameter.

The distribution of $W$ admits a normal-gamma mixture representation. Its unconditional density can therefore be obtained by integrating the conditional Gaussian density with respect to the distribution of $\gamma$. 
We begin by deriving the unconditional moments of the increments in $W$.

\begin{prop}[Unconditional moments of $W(t)$]
	For any $n \in \mathbb{N}$, the $n$-th unconditional raw moment of the increments in $W$ is
	\begin{align}
		\E \left[ \left( W(t + h) - W(t)  \right)^n \right]
		&= \sum_{k=0}^{\lfloor n/2  \rfloor} \frac{n!}{2^k k! (n-2k)!} \theta^{n-2k}  \sigma^{2k} 
		v^{n - 2k(1-H)} \frac{\Gamma \left(\frac{h}{v} + m-2k(1-H)\right)}{\Gamma \left(\frac{h}{v} \right)},
		\label{eqn: fVG model, unconditional raw moments of W(t)}
	\end{align}
	and the $n$-th unconditional central moment of the increments in $W$ is
	\begin{align}
		&\E \left[ \left( W(t + h) - W(t)  - \E[ W(t+h) - W(t) ] \right)^n \right] \nonumber  \\
		&\qquad = \sum_{k=0}^{\lfloor n/2  \rfloor}  \frac{n!}{2^k k! (n-2k)!}   \theta^{n-2k}  \sigma^{2k}  \sum_{\ell=0}^{n-2k} \binom{n-2k}{\ell} (-h)^{n-2k -\ell}  v^{2kH + \ell} \frac{\Gamma \left(\frac{h}{v} + 2kH + \ell \right)}{\Gamma \left(\frac{h}{v}\right)} .
		\label{eqn: fVG model, unconditional central moments of X(t)}
	\end{align}
	\label{prop: fVG model, unconditional moments of W(t)}
\end{prop}

Following Proposition~\ref{prop: fVG model, unconditional moments of W(t)}, explicit expressions for the first four moments are presented below. These expressions are subsequently used in Section~\ref{sec: Parameter Estimation} for parameter estimation.

\begin{coro}
	For any $t \geq 0$ and $h \geq 0 $, the unconditional expectation and the second through fourth unconditional central moments of the increments in $W$ are
	\begin{align*}
		\E \left[W(t+h) - W(t) \right]
		&= \theta h \\
		\E \left[ \left( W(t + h) - W(t) - \E[ W(t+h) - W(t) ] \right)^2 \right]
		&= \theta^2 v h   +    \sigma^2 v^{2H} \frac{\Gamma \left(\frac{h}{v} + 2H \right)}{\Gamma \left(\frac{h}{v} \right)} \\
		\E \left[ \left( W(t + h) - W(t) - \E[ W(t+h) - W(t) ] \right)^3 \right]
		&=   2 \theta^3 v^2 h + 6 H \sigma^2 \theta v^{1+2H} \frac{\Gamma\left(\frac{h}{v} + 2H \right) }{\Gamma\left(\frac{h}{v} \right)}   \\
		\E \left[ \left( W(t + h) - W(t) - \E[ W(t+h) - W(t) ] \right)^4 \right] \nonumber \\
		&\hspace{-7cm} =   \theta^4  (3 h^2 v^2 + 6 h v^3)
		+ 6  \sigma^2 \theta^2 v^{2H+1} (h + 2H (2H+1)v ) \frac{\Gamma\left(\frac{h}{v} + 2H \right) }{\Gamma\left(\frac{h}{v} \right)} 
		+ 3 \sigma^4 v^{4H} \frac{\Gamma\left(\frac{h}{v} + 4H \right) }{\Gamma\left(\frac{h}{v} \right)}.
	\end{align*}
	\label{coro: fVG model, first four unconditional central moments of X(t)}
\end{coro}

\begin{figure}[!htp]
	\centering
	\includegraphics[width = .95\linewidth]{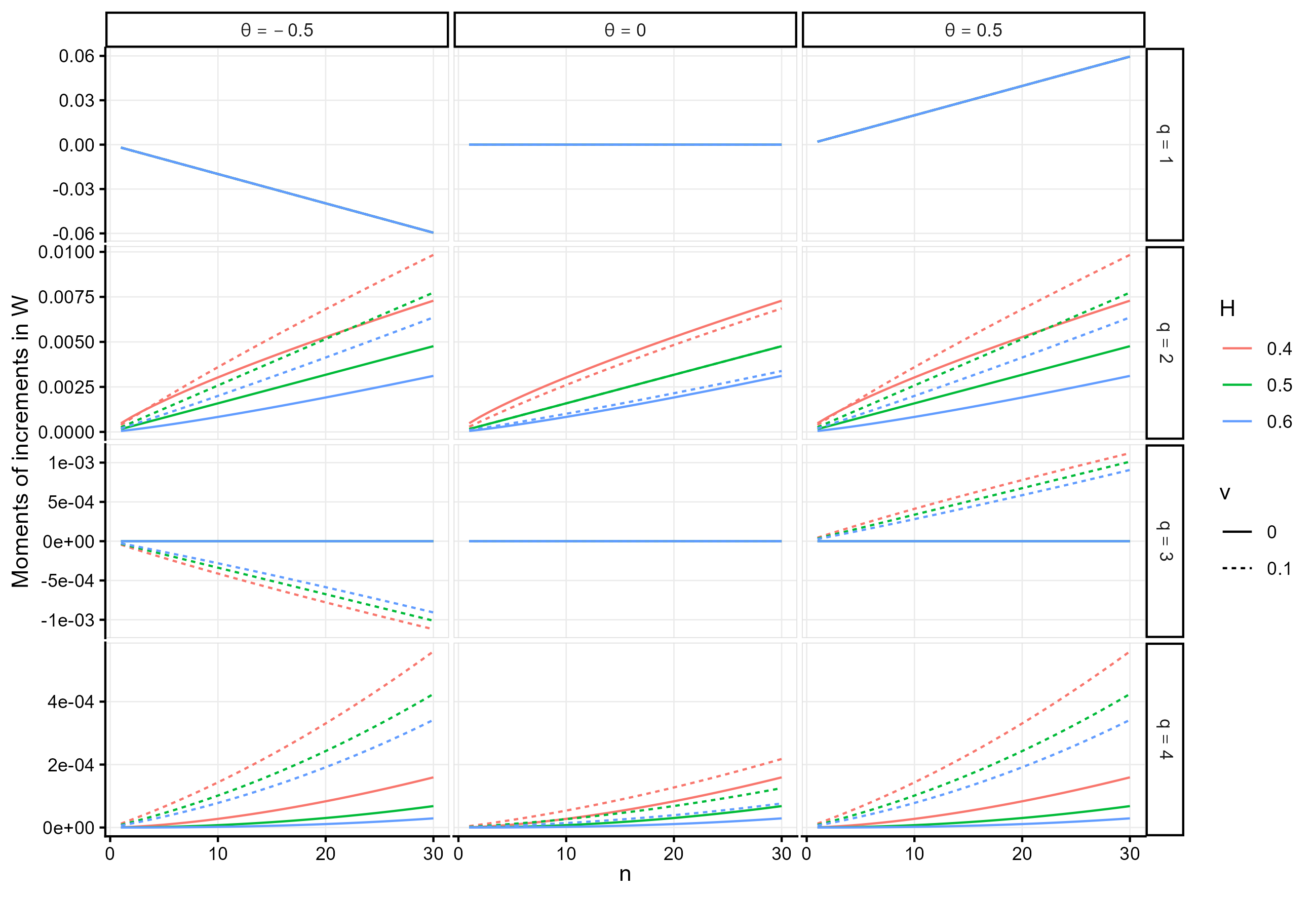}
	\caption{Moments of the increments in $W$ given in Corollary \ref{coro: fVG model, first four unconditional central moments of X(t)} across parameters $\{H,v,\theta\}$. }
	\label{fig: time changed fBm, Moments of W}
\end{figure}

Figure~\ref{fig: time changed fBm, Moments of W} illustrates the moments reported in Corollary~\ref{coro: fVG model, first four unconditional central moments of X(t)} for time lags $h = n/252$, where $n \in \{1,\dots,30\}$. Under this convention, $n$ may be interpreted as the number of trading days. Throughout, we fix $\sigma = 0.2$, as this parameter primarily affects the scale of $W$ and does not materially alter the qualitative behavior of its moments. 

When $\theta=0$, both the first raw moment and the third central moment remain identically zero, indicating that $\theta$ is the primary driver of the mean and skewness of $W$. As $\theta$ increases, the skewness becomes more pronounced, with the effect being particularly strong for smaller values of $H$.
The Hurst parameter $H$ also influences the rate at which higher-order moments evolve with the time lag $h$. In particular, using the asymptotic relation $\Gamma(h/v+2kH)/\Gamma(h/v) \sim (h/v)^{2kH}$ as $h \rightarrow \infty$, the central moments of $W$ grow according to a power law depending on the value of $H$.

Similar to Section \ref{ssec: time changed fBm, Properties}, the parameter $v$ governs the tail heaviness, or kurtosis, of $W$.
When $\theta=0$, the kurtosis is monotonically increasing in $v$ and converges to $3$ as $v \rightarrow 0$.
Since $v$ governs the variability of the activity-time process, larger values of $v$ correspond to greater fluctuations in the gamma activity clock, which in turn induces larger fluctuations in the conditional variance and produces heavier tails of $W$.

Overall, Figure~\ref{fig: time changed fBm, Moments of W} demonstrates that $W$ provides separate control over the key distributional features of financial returns, allowing it to simultaneously capture skewness, heavy tails, and long-range dependence, which are three prominent features of financial return data.

\section{The Fractional Variance Gamma (fVG) Model}
\label{sec: fVG model}

In this section, we use the preceding results concerning time-changed fBm to construct the fVG model within the standard continuous-time, continuous trading arbitrage-free option pricing framework. The key idea is that the randomness in an asset price process, represented by a generic pure-jump innovation process $Y$, is generated by a time-changed fBm, $X=B_{H}(\gamma)$, where the fBm $B_{H}$ is evaluated at gamma activity time $\gamma$.

The adoption of activity time has a long tradition in option pricing and has a clear economic interpretation; see, e.g., \citet{MadanCarrChang1998VGoption, GemanEtAl2001TimeChangesLevy, MadanWang2023EconomicsTime}. Activity time corresponds to the times at which trades are executed and therefore reflects the accumulation of trading activity. During periods of high trading intensity, prices are updated more frequently and typically exhibit greater volatility. By contrast, between trading times, there are no transactions and hence no observable changes in prices. Consequently, for option valuation it is natural to model the price process in activity time, which reflects the flow of trades, rather than calendar time, which evolves exogenously at a constant rate.

\subsection{The Setup}
\label{ssec: fVG model, The Setup}

Consider a frictionless and competitive market consisting of a stock and a default-free money market account (MMA) that trade continuously over the time horizon $[0,T]$. Without loss of generality, we assume that the stock pays no cash flows and that the risk-free spot rate process $r=(r(t))_{t\in[0,T]}$ is deterministic. Dividend payments can be incorporated at the cost of additional notation, and stochastic interest-rate dynamics can be introduced through standard extensions; neither affects the key contribution of the paper.

The fractional Variance Gamma (fVG) model is characterized by the stock price process
\begin{align}
	S(t) &= S(0)\exp\left\{ \Xi(t) + Y(t) \right\} ,
	\label{eqn: Equity model, fVG stock price under P} 
\end{align}
where $S(0)>0$ and $\Xi=(\Xi(t))_{t\in[0,T]}$ denotes a predictable drift process. For convenience, we assume that $\Xi$ is absolutely continuous, so that
\begin{equation*}
	\Xi(t)=\int_{0}^{t}\xi(s)\diff s 
\end{equation*}
for all $t\in[0,T]$, where $\xi$ is a predictable, locally integrable process. Accordingly, $\Xi$ is continuous with finite variation. 
The process $Y = (Y(t))_{t\in[0,T]}$ represents the stock's innovation process and is modeled as a pure-jump process with jump measure $\mu_Y$. Let $\nu_Y$ denote its compensator with density $\psi_Y$ under the statistical measure $\mathbb{P}$. Following Section~\ref{ssec: Stochastic Processes, Marked point process}, the processes $Y$ and its compensator process $A_Y$ under $\mathbb{P}$ admits the representations
\begin{align*}
	Y(t) = \int_0^t \int_\mathbb{R} x \, \mu_Y(\diff s, \diff x), \qquad
	A_Y(t) = \int_0^t \int_\mathbb{R} x \, \nu_Y(\diff s, \diff x) \qquad 
	\forall \ t \in [0,T].
\end{align*}

Although the valuation results developed below apply to any pure-jump innovation process, our primary interest is in innovation $Y$ generated from the time-changed fBm $X$. For example, $Y$ may be constructed by enriching the jump structure of $X$ through an arbitrary predictable intensity function $\eta: \mathbb{R}_+ \times \mathbb{R} \rightarrow \mathbb{R}_+$, yielding 
\begin{align*}
	Y(t) = \int_0^t \int_\mathbb{R} \eta(s,x) \, \mu_X(\diff s, \diff x) ,
\end{align*} 
where $\mu_X$ is the jump measure of the time-changed fBm $X$. Alternatively, following the VG construction of \citet{MadanCarrChang1998VGoption}, one may take the innovation process to be
\begin{align*}
	Y(t) = W(t) = \theta \gamma(t) + \sigma B_H(\gamma(t)) ,
\end{align*}
introduced in Section~\ref{ssec: time changed fBm, Skewness}. Since $W$ is likewise a pure-jump process, the valuation framework developed here applies directly to this specification.

Taking logarithms of Equation \eqref{eqn: Equity model, fVG stock price under P}, the log price process is linear in the innovation $Y$, 
\begin{align}
	\ln S(t) 
	& =\ln S(0)+ \Xi(t) +  Y(t) . 
	\label{eqn: Equity model, fVG log stock returns process under P} 
\end{align}
A lower value of the Hurst exponent $H$ produces rougher stock price trajectories, while a higher variance intensity $v$ makes the stock price process less regular, exhibiting more pronounced periods of inactivity interspersed with abrupt jumps, characterized by extended flat segments punctuated by sudden spikes.

Under the statistical probability measure $\mathbb{P}$, the conditional expected log return of the stock over the interval $[u,t]$, given the information set $\mathcal{F}(u)$, is 
\begin{align}
	\E^\mathbb{P}\left[\ln \frac{S(t)}{S(u)} \, \middle| \, \mathcal{F}(u-)\right]
	&=\bigg( \E^\mathbb{P}\left[  \Xi(t) \, \middle| \, \mathcal{F}(u-)\right]    - \Xi(u)  \bigg)
	+  \bigg( \E^\mathbb{P}\left[ A_Y(t) \, \middle| \, \mathcal{F}(u-)\right]  -  A_Y(u)  \bigg) ,
	\label{eqn: Equity model, fVG conditional expected stock returns under P}
\end{align} 
where the compensator process $A_Y$ depends on the latent activity time $\gamma$. 
The first term represents the contribution of the exogenous drift process $\Xi$, whereas the second captures the endogenous drift induced by the dependence structure of the innovation process $Y$. When $H = 0.5$, $Y$ reduces to a time-changed Brownian motion with independent increments, so the endogenous component vanishes and expected log returns is determined entirely by the exogenous drift.

More interestingly, when $H \neq 0.5$, log stock returns exhibit serial dependence arising from the endogenous drift component, characterized by a power-law decay of autocorrelations.
For $H > 0.5$, log returns exhibit positive autocorrelation and long-range dependence, implying that periods of positive historical returns tend to be followed by higher expected future returns. Moreover, the influence of past increments decays slowly and remains significant even over long horizons.   
Conversely, when $H < 0.5$, increments are negatively correlated, giving rise to anti-persistent dynamics. In this regime, positive returns are more likely to be followed by negative returns and vice versa.

The conditional variance of the stock's log returns is given by
\begin{align} 
	\Var^\mathbb{P}\left[\ln \frac{S(t)}{S(u)} \middle| \mathcal{F}(u-)\right]
	&=\Var\bigg[  \Xi(t)-\Xi(u)  + Y(t)- Y(u)  \bigg| \mathcal{F}(u-)\bigg] .
	\label{eqn: Equity model, fVG conditional variance stock returns under P} 
\end{align} 
In many applications, the innovation process $Y$ is assumed to be the sole source of randomness, with $\Xi$ taken to be deterministic. Under this assumption, the conditional variance simplifies to the expected quadratic variation of the jump process $Y$, 
\begin{align*}
	\Var^\mathbb{P}\left[\ln \frac{S(t)}{S(u)} \middle| \mathcal{F}(u-)\right]
	=\Var^\mathbb{P}\left[Y(t)-Y(u)  \middle| \mathcal{F}(u-)\right]
	= \E^\mathbb{P} \left[ \int_u^t \int_\mathbb{R} x^2 \, \nu_Y(\diff s, \diff x) \middle| \mathcal{F}(u) \right] .
\end{align*}
Importantly, the Hurst exponent $H$ governs the temporal scaling of the variance.
When $H > 0.5$, log returns are superdiffusive, with the variance growing faster than linearly over time. Conversely, when $H < 0.5$, log returns are subdiffusive, with the variance increasing at a rate slower than linear in time.

\begin{remark}[Continuous Brownian motion]
	The model can be easily extended to include a continuous Brownian component. In this setting, the Itô formula for general semimartingales naturally accommodates the presence of a continuous martingale term alongside the jump component.
	Nevertheless, such an extension is not pursued in the present study. Since the time-changed fBm generates an infinite-activity process, its sample paths are, in practice, virtually indistinguishable from those of a continuous process when observed at discrete time intervals. This observation is analogous to the argument contained in \citet{MadanCarrChang1998VGoption} in the context of the VG model. 	
	To maintain a concise exposition and focus on the principal contributions of this paper, we omit the additional source of randomness arising from additional Brownian motions.
\end{remark}

\subsection{No Arbitrage and Existence of an EMM}

Let $\bar{S} = (\bar{S}(t))_{t \in [0,T]} = (e^{-\int_0^t r(c) \diff c} S(t))_{t \in [0,T]}$ denote the discounted stock price process, and let $\mathbb{Q}$ denote the equivalent martingale measure (EMM), defined as the probability measure $\mathbb{Q} \sim \mathbb{P}$ under which $\bar{S}$ is a $\mathbb{Q}$-martingale under the filtration $\mathcal{F}$. The absence of arbitrage imposes a restriction on the drift of the stock price process under the statistical measure $\mathbb{P}$, as characterized in Proposition~\ref{prop: Equity model, Stock price dynamics under Q} below.

\begin{prop}[Stock price dynamics under EMM]
	Suppose there exists an EMM $\mathbb{Q}$ whose compensator satisfies $\tilde{\nu}_Y(\diff t, \diff x) = \lambda(t,x)  \nu_Y(\diff t, \diff x)$ for all $t\in[0,T],x\in\mathbb{R}$, and for some predictable process $\lambda=(\lambda(t,x))_{t\in[0,T],x\in\mathbb{R}}$. 
	Then, the stock's drift must satisfy the arbitrage-free drift condition
	\begin{align}
		r(t) = \xi(t) + \int_\mathbb{R} (e^x-1) \, \lambda(t,x) \, \psi_Y(t,x) \, \diff x ,  \qquad \forall \, t \in [0,T],
		\label{eqn: Equity model, fVG arbitrage free drift condition}
	\end{align}
	where $\lambda$ denotes the market price of jump risk.
	Furthermore, under $\mathbb{Q}$, the discounted stock price $\bar{S}$ satisfy
	\begin{align}
		\bar{S}(t) 
		= \bar{S}(u)  + \int_u^t \int_\mathbb{R} x \, \bar{S}(c-)\left(e^{x}-1\right) \, (\mu_Y-\tilde{\nu}_Y)(\diff c,\diff x) ,
		\qquad 0 \leq u \leq t \leq T,
		\label{eqn: fVG model, Stock price under Q}
	\end{align}
	where $\tilde{\nu}_Y(\diff t, \diff x) = \lambda(t,x) \, \nu_Y(\diff t, \diff x)$.
	\label{prop: Equity model, Stock price dynamics under Q}
\end{prop}

From Equation \eqref{eqn: fVG model, Stock price under Q}, the conditional variance of the discounted stock price process is  
\begin{align}
	\Var^\mathbb{Q}\left[ \bar{S}(t) \,\middle|\, \mathcal{F}(u) \right]
	= \E^\mathbb{Q}\left[ \int_u^t \int_\mathbb{R} x^2 \, \bar{S}(c-)^2 \left(e^{x}-1\right)^2 \tilde{\nu}_Y(\diff c, \diff x) \, \middle| \, \mathcal{F}(u) \right] .
\end{align}  
By substituting the arbitrage-free drift condition in Equation \eqref{eqn: Equity model, fVG arbitrage free drift condition} into the stock price dynamics, the drift process $\Xi$ can be endogenized, and the stock's log returns can be expressed as 
\begin{align}
	\ln \frac{S(t)}{S(u)}
	&=
	\underbrace{\int_u^t r(c)\,\diff c}_{\text{\shortstack{Cumulative\\risk-free rate}}}
	-
	\underbrace{\int_u^t \int_\mathbb{R} (e^x-1)\lambda(t,x)\psi_X(t,x)\,\diff x\,\diff c}_{\text{Risk premium}}
	\ +
	\underbrace{\vphantom{\int_u^t} A_Y(t) - A_Y(u)}_{\text{\shortstack{Endogenous\\drift}}} \nonumber \\
	&\qquad + 
	\underbrace{\vphantom{\int_u^t} Y(t) - Y(u) -(A_Y(t) - A_Y(u))}_{\text{Martingale}}.
	\label{eqn: Equity model, fVG model, Decomposition of stock returns into 4 components}
\end{align}
The log return process therefore admits a decomposition into four distinct components: (i) the cumulative risk-free return, (ii) a risk premium determined by the market price of jump risk $\lambda$, (iii) an endogenous drift arising from the dependence structure and historical trajectory of the innovation process $Y$, and (iv) a martingale component capturing the residual uncertainty in stock returns.

While Proposition~\ref{prop: Equity model, Stock price dynamics under Q} yields the necessary condition on the drift of the stock when an EMM exists, the following result provides a partial converse to Proposition~\ref{prop: Equity model, Stock price dynamics under Q} by establishing sufficient conditions for the existence of an equivalent martingale measure.

\begin{prop}[Sufficient conditions for the existence of EMM]
	Suppose there exists a predictable market price of jump risk process $\lambda=(\lambda(t,x))_{t\in[0,T],x\in\mathbb{R}}$ satisfying
	\begin{enumerate}
		\item Strict positivity: $\lambda(t,x) > 0$ for all $t \in [0,T]$ and $x \in \mathbb{R}$,
		\item Novikov's condition: $\E^\mathbb{P} \left[ \exp\left( \frac{1}{2} \int_0^T \int_\mathbb{R} (\lambda(c,x) - 1)^2 \, \nu_Y(\diff c, \diff x)  \right) \right] < \infty$, 
		\item True martingale condition: $\E^\mathbb{P} \left[ \int_0^T \int_\mathbb{R} \, x^2 \, \lambda(c,x) \, \nu_Y(\diff c, \diff x)  \, \right] < \infty$, and
		\item Arbitrage-free drift condition: $r(t) = \xi(t) + \int_\mathbb{R} (e^x-1) \, \lambda(t,x) \, \psi_X(t,x) \, \diff x$ for all $t \in [0,T]$.
	\end{enumerate} 
	Then, there exists an EMM $\mathbb{Q}$.
	\label{prop: Sufficient conditions for the existence of EMM}
\end{prop}

However, the presence of jumps in the stock price dynamics generally leads to market incompleteness. An implication of this proposition is that the EMM is generally non-unique, with the set of admissible EMMs consisting of equivalent probability measures $\mathbb{Q}(\lambda)$ corresponding to market prices of jump risk $\lambda$ that satisfy the four stated hypotheses.

\subsection{Derivative Valuation}
\label{ssec: fVG model, Derivative Valuation}

For valuation purposes, we work within an arbitrage-free market. By arbitrage-free, we mean that the market satisfies no free lunch with vanishing risk (NFLVR) and no dominance (ND). Under the fundamental theorems of asset pricing, these conditions imply the existence of an equivalent martingale measure (EMM) $\mathbb{Q}$. Specifically, we assume that the stock price process under the fVG model satisfies the sufficient conditions for the existence of an EMM established in Proposition~\ref{prop: Sufficient conditions for the existence of EMM}.

However, the presence of jumps generally leads to market incompleteness. By the second fundamental theorem of asset pricing, the non-uniqueness of the EMM implies that the fVG market is incomplete. Therefore, derivative valuation requires the selection of an EMM $\mathbb{Q}(\lambda)$ from the admissible set. Several approaches have been proposed in the literature for selecting an EMM. One approach determines a unique derivative price through an optimization criterion, such as utility indifference pricing. A second approach selects a unique EMM based on economic considerations, such as the treatment of diversifiable risk, or analytical criteria, such as the minimal martingale measure. A third approach empirically identifies an EMM consistent with observed prices of traded derivatives.

Once an EMM $\mathbb{Q}$ has been selected through any of these approaches or otherwise, the arbitrage-free price at time $t$ of a derivative with payoff function $G:\mathbb{R}\to\mathbb{R}$ and maturity $T$ is given by the risk-neutral valuation formula 
\begin{align*}
	\E^{\mathbb{Q}}\left[e^{-\int_{t}^{T} r(u)\diff u} \, G\left(S(T)\right)\,\middle|\,\mathcal{F}(t)\right] .
\end{align*}
This expectation can be evaluated numerically using the conditional distribution simulation methodology developed in Section~\ref{ssec: Time-changed fBm process, Semimartingale Behavior and Equivalent Probability Measure Changes}.


To simplify the model and improve analytical tractability, one may impose that the market price of jump risk is independent of jump size and varies only with time. Specifically, assume that
\begin{align}
	\lambda(t,x) \equiv \lambda(t) > 0.
	\label{eqn: Equity model, fVG simplifying assumption of lambda}
\end{align}
Economically, this restriction implies that the change of measure adjusts the jump compensator proportionally across all jump sizes, so that positive and negative jumps of different magnitudes are subject to the same relative risk adjustment at a given point in time.
In this case, since $\lambda$ no longer depends on $x$, it can be factored out of the integral in the arbitrage-free drift condition in Equation~\eqref{eqn: Equity model, fVG arbitrage free drift condition}, yielding the explicit representation
\begin{align*}
	\lambda(t) &= -\frac{\xi(t) - r(t)}{\int_\mathbb{R} (e^x-1)   \psi_Y(t,x) \diff x}.
\end{align*}
Consequently, once the physical dynamics of the stock price have been estimated, including the drift $\xi$ and jump compensator $\psi_Y$, the market price of jump risk $\lambda$ and the corresponding EMM $\mathbb{Q}$ are determined without requiring an independent calibration of the risk-neutral measure.

Specifically, let $L = (L(t))_{t \in [0,T]}$ denote the likelihood process associated with the change of measure from $\mathbb{P}$ to $\mathbb{Q}$. Substituting the expression for the market price of jump risk into the Girsanov density yields
\begin{align*}
	L(t) 
	&=\exp\bigg\{ \int_0^t \int_\mathbb{R} \left(\frac{\xi(s) - r(s)}{\int_\mathbb{R} (e^x-1)   \psi_Y(s,x) \diff x} + 1\right) \, \nu_Y(\diff s, \diff x) \\
	&\qquad\qquad + \int_0^t \int_\mathbb{R} \log \left( \frac{\xi(s) - r(s)}{\int_\mathbb{R} (e^x-1)   \psi_Y(s,x) \diff x} \right) \, \mu_Y(\diff s,\diff x)   \bigg\} .
\end{align*} 
Then, by Bayes' theorem for conditional expectations, the arbitrage-free price of a contingent claim with payoff $G(S(T))$ at maturity $T$ can be written as
\begin{align*}
	\E^{\mathbb{Q}}\left[e^{-\int_{t}^{T} r(u)\diff u} \, G\left(S(T)\right) \,\middle|\,\mathcal{F}(t)\right] 
	= \frac{1}{L(t)}\E^{\mathbb{P}}\left[e^{-\int_{t}^{T} r(u)\diff u} \, G\left(S(T)\right) L(T) \,\middle|\,\mathcal{F}(t)\right] ,
\end{align*}
Accordingly, derivative valuation can be performed under the statistical measure $\mathbb{P}$ by evaluating the conditional distribution of the innovation process $Y$ together with the likelihood adjustment induced by the change of measure $L(T)/L(t)$. Following the simulation methodology discussed in Section~\ref{ssec: Time-changed fBm process, Semimartingale Behavior and Equivalent Probability Measure Changes}, the arbitrage-free price can therefore be obtained through simulations of the conditional distribution of $Y$ under $\mathbb{P}$.

\subsection{Alternative time-changed fgBm model}

In the preceding section, stock prices were specified as an exponential transformation of the innovation process $Y$, yielding a log return process that is linear in $Y$. Specifically, the fVG model was constructed by specifying
\begin{align*}
	\diff \log S(t) = \xi(t) \diff t + \diff Y(t). 
\end{align*}
Motivated by the classical Black--Scholes--Merton paradigm, an alternative and equally natural specification is to generalize geometric Brownian motion by replacing standard Brownian motion with the time-changed fBm process. This section considers this alternative formulation and demonstrates that the resulting model can be represented within the same valuation framework through an appropriate transformation of the jump intensity.

The time-changed fractional geometric Brownian motion (tfgBm) model is defined by
\begin{align*}
	d S(t) &=  S(t) \xi(t) \diff t +  S(t-) \diff Y(t) ,
\end{align*} 
where $S(0) > 0$ is given. To ensure the positivity of the stock price process, we impose the condition that $\Delta Y(t) > -1$ for all $t \in [0,T]$.
The corresponding solution $S$ is given by the Doléans--Dade exponential,
\begin{align}
	S(t) 
	&= S(0) \, \exp\left\{ \, \Xi(t)  \right\} \, \prod_{u \in [0,t]} (1+ \Delta Y(u)) \nonumber \\ 
	&= S(0)\exp\left\{ \Xi(t) + \int_0^t \int_\mathbb{R} \ln(1+ x) \, \mu_Y(\diff u,\diff x) \right\}. 
	\label{eqn: Equity model, tfgBm stock price process}
\end{align}  
Accordingly, the log return process admits the representation
\begin{align*}
	\ln \frac{S(t)}{S(u)} 
	&= \int_u^t \xi(c) \diff c
	+ \int_u^t \int_\mathbb{R} \ln(1+ x) \, \nu_Y(\diff c, \diff x) 
	+ \int_u^t \int_\mathbb{R} \ln(1+ x) \, (\mu_Y-\nu_Y)(\diff c, \diff x) .
\end{align*}
The first term represents the exogenous drift, while the second term is the endogenous predictable drift induced by the innovation process $Y$. The final term is a $\mathbb{P}$-martingale that captures the residual randomness in the log return dynamics.

Because the tfgBm framework is formulated directly through the stock price SDE, the derivation of the no-arbitrage drift condition is comparatively straightforward. Suppose there exists an EMM $\mathbb{Q}$. 
Then, under $\mathbb{Q}$, the compensated measure $\mu_Y - \tilde{\nu}_Y$ is a $\mathbb{Q}$-martingale with $\tilde{\nu}_Y(\diff t, \diff x) = \lambda(t,x) \, \nu_Y(\diff t, \diff x)$, and the discounted stock price process $\bar{S}$ satisfies
\begin{align*}
	\bar{S}(t) 
	&= \bar{S}(u) + \int_0^t \bar{S}(c)  \left[ \xi(c) - r(c) + \int_\mathbb{R} x \, \lambda(c,x) \, \psi_Y(c,x) \diff x \right]  \diff c 
	+ \int_0^t \int_\mathbb{R} x \, \bar{S}(c-)   \, (\mu_Y-\tilde{\nu}_Y)(\diff c, \diff x) .
\end{align*} 
Thus, the arbitrage-free drift condition is
\begin{align}
	r(t) = \xi(t) + \int_\mathbb{R} x \, \lambda(t,x) \,  \psi_Y(t,x) \, \diff x,
	\qquad \forall \, t \in [0,T].
	\label{eqn: Equity model, tfgBm arbitrage free drift condition}
\end{align}
Similar to the fVG framework, the no-arbitrage condition yields an economically meaningful decomposition of the stock's log returns. In particular, $\ln (S(t)/S(u))$ can be expressed as the sum of four components: (i) the cumulative risk-free return $\int_u^t r(c)\,\diff c$, (ii) the compensation for jump risk $\int_u^t \int_\mathbb{R} x \, \lambda(t,x)\psi_X(t,x)\,\diff x\,\diff c$, (iii) an endogenous drift arising from past realizations $\int_u^t \int_\mathbb{R} \ln(1+ x) \, \nu_Y(\diff c,\diff x)$, and (iv) residual martingale innovation process $\int_u^t \int_\mathbb{R} \ln(1+ x) \, (\mu_Y - \nu_Y)(\diff c,\diff x)$. This decomposition is the direct analogue of Equation~\eqref{eqn: Equity model, fVG model, Decomposition of stock returns into 4 components}, with the principal distinction being the jump-amplitude transformation $x \mapsto \ln(1+x)$.

The price of a contingent claim with terminal payoff $G(S(T))$ remains 
$
	E^{\mathbb{Q}}[e^{-\int_{t}^{T}r(v)\diff v} \, G(S(T)) |\mathcal{F}(t)],
$
where the terminal stock price under $\mathbb{Q}$ admits the representation 
\begin{align*}
	\ln \frac{S(t)}{S(u)}
	&=
	\int_u^t r(c)\,\diff c
	-
	\int_u^t \int_\mathbb{R} \left( x - \ln(1+ x) \right)  \lambda(t,x) \psi_X(t,x)\,\diff x\,\diff c
	+\int_u^t \int_\mathbb{R} \ln(1+ x) \, (\mu_Y - \tilde{\nu}_Y)(\diff c,\diff x). 
\end{align*}
and 
$\lambda(t,x)$ is determined by the no-arbitrage restriction in Equation~\eqref{eqn: Equity model, tfgBm arbitrage free drift condition}.
As in the fVG framework, additional tractability is obtained by assuming that the market price of jump risk is independent of jump size. Under this restriction, the market price of risk simplifies to
\begin{align}
	\lambda(t,x) &= -\frac{\xi(t) - r(t)}{\int_\mathbb{R} x \, \psi_Y(t,x) \diff x},
\end{align} 
and risk-neutral valuation can be likewise performed without requiring a separate calibration of the market price of risk.

A direct comparison of Equations~\eqref{eqn: Equity model, fVG stock price under P} and~\eqref{eqn: Equity model, tfgBm stock price process} reveals that the fVG and tfgBm specifications are equivalent up to a deterministic transformation of the jump amplitudes.
Let 
\begin{align*}
	Y_2(t) =  \int_0^t \int_\mathbb{R} x \, \mu_{Y_2}(\diff s, \diff x)
\end{align*}
denote the innovation process in the tfgBm model. Then, the same stock price dynamics can be obtained under the fVG framework by adopting the innovation process
\begin{align*}
	Y_1(t) = \int_0^t \int_\mathbb{R} \ln(1+x) \, \mu_{Y_2}(\diff s, \diff x).
\end{align*}
That is, the fVG innovation process $Y_1$ is obtained by applying the jump-amplitude transformation  
\begin{align*}
	\eta(x) := \ln(1+x)
\end{align*}
to the jumps of the tfgBm innovation process $Y_2$.
Conversely, given an fVG innovation process
$
	Y_1(t) = \int_0^t \int_\mathbb{R} x \, \mu_{Y_1}(\diff s, \diff x),
$
the corresponding tfgBm innovation process is obtained by applying the inverse jump-amplitude transformation
\begin{align*}
	\eta^{-1}(x) = e^x - 1.
\end{align*}

Consequently, the principal results developed in Sections~\ref{ssec: fVG model, The Setup} through \ref{ssec: fVG model, Derivative Valuation} extend directly to the tfgBm framework after  with the appropriate modification in the jump amplitude of the innovation process. Of the two specifications, the fVG model offers greater analytical tractability. Since log returns are linear in the driving innovation process $Y$, their statistical properties can be characterized more directly, facilitating both theoretical analysis and economic interpretation. By contrast, under the tfgBm specification, log returns depend on the modified jump amplitudes through the mapping $x \mapsto \ln(1+x)$, which complicates the characterization of return dynamics and the derivation of explicit pricing results.

\section{Parameter Estimation via GMM}
\label{sec: Parameter Estimation}

This section presents a Generalized Method of Moments (GMM) procedure for estimating the parameters of the fVG model. We consider the constant-parameter stock price process 
\begin{align*}
	S(t) 
	= S(0) \exp\left\{ \xi t + \theta \gamma(t; v) + \sigma B_H(\gamma(t; v)) \right\},
\end{align*}
where $\xi \in \mathbb{R}$, $\theta \in \mathbb{R}$, $\sigma \geq 0$, $v \geq 0$, and $H \in (0,1)$; see Section~\ref{ssec: time changed fBm, Skewness}. Several well-known models arise as special cases of the proposed evolution. When $H=0.5$, the innovation process reduces to standard Brownian motion, and the model coincides with the classical VG model of \citet{MadanCarrChang1998VGoption}. Setting $v=0$ and $\theta=0$ eliminates the stochastic time-change, yielding the fractional Black--Scholes--Merton (BSM) model driven by fBm. Although this model is known to admit arbitrage opportunities, it is included for completeness and as a benchmark for the full specification. Finally, when $H=0.5$, $\theta=0$, and $v=0$, the model further simplifies to the classical BSM framework.

Suppose stock prices are observed discretely as $\{S_{t_i}\}_{i=0}^N$ on the time grid $\{t_i\}_{i=0}^N$. For simplicity, we assume equally spaced observations with sampling interval $d>0$, so that $t_i = i d$ for all $i \in {0,1,\dots,N}$. Let $R_i^{(n)}$ denote the $i$-th observation of the $n$-lag log returns,  
\begin{align*}
	R_i^{(n)} := \ln \left(S_{t_{i+n}} / S_{t_{i}} \right),
\end{align*}
where $n \in \mathbb{N}$, so that each lag corresponds to the time interval $nd$.
Following Corollary~\ref{coro: fVG model, first four unconditional central moments of X(t)}, let $\rho(q,n)$ denote the unconditional mean (for $q=1$) and central moments (for $q>1$) of the $n$-lag log returns. Specifically,  
\begin{align*}
	\rho(q,n) 
	&= 
	\begin{cases}
		\E\left[ R_i^{(n)}  \right], & q=1 \\
		\E\left[\left(R_i^{(n)} - \E\left[R_i^{(n)}\right]\right)^q\right] , & q \in \{2,3,4\}
	\end{cases} \\ 
	&= 
	\begin{cases}
		(\xi + \theta) nd, & q=1 \\
		\theta^2 v nd   +    \sigma^2 v^{2H} \frac{\Gamma \left(\frac{nd}{v} + 2H \right)}{\Gamma \left(\frac{nd}{v} \right)} , & q=2 \\
		2 \theta^3 v^2 nd + 6 H \sigma^2 \theta v^{1+2H} \frac{\Gamma\left(\frac{nd}{v} + 2H \right) }{\Gamma\left(\frac{nd}{v} \right)} , & q=3  \\
		 \theta^4  (3 (nd)^2 v^2 + 6 nd v^3)
		+ 6  \sigma^2 \theta^2 v^{2H+1} (nd + 2H (2H+1)v ) \frac{\Gamma\left(\frac{nd}{v} + 2H \right) }{\Gamma\left(\frac{nd}{v} \right)} \\
		\qquad+ 3 \sigma^4 v^{4H} \frac{\Gamma\left(\frac{nd}{v} + 4H \right) }{\Gamma\left(\frac{nd}{v} \right)}, & q=4 .
	\end{cases} 
\end{align*}
The corresponding sample moments are defined by
\begin{align}
	\hat{\rho}(q,n) 
	&= \begin{cases}
		\frac{1}{N-n+1} \sum_{i=0}^{N-n} R_i^{(n)}, & q =1 \\
		\frac{1}{N-n+1} \sum_{i=0}^{N-n} \left(R_i^{(n)} - \frac{1}{N-n+1} \sum_{j=0}^{N-n} R_j^{(n)} \right)^q, & q \in \{2,3,4\}.
	\end{cases}
\end{align}

Let $n_q \in \mathbb{N}$ denote the number of moment conditions of order $q$, and define $L = n_1 + n_2 + n_3 + n_4$. We then construct the $(L \times 1)$ vector of moment conditions as the stacked differences between theoretical and empirical moments,
\begin{align*}
	\mathbf{m} = \big[
		&\rho(1,1) - \hat{\rho}(1,1), \quad  
		\dots  \quad
		\rho(1,n_1) - \hat{\rho}(1,n_1),  \quad
		\rho(2,1) - \hat{\rho}(2,1), \quad
		\dots \quad
		\rho(2,n_2) - \hat{\rho}(2,n_2), \\
		&\rho(3,1) - \hat{\rho}(3,1), \quad
		\dots \quad 
		\rho(3,n_3) - \hat{\rho}(3,n_3), \quad
		\rho(4,1) - \hat{\rho}(4,1), \quad
		\dots \quad 
		\rho(4,n_4) - \hat{\rho}(4,n_4)
	\big]^T.
\end{align*}

For a given $(L \times L)$ weighting matrix $\mathbf{W}$, the GMM estimator is defined as
\begin{align}
	\{\xi^*, \theta^*, v^*,H^*,\sigma^*\}
	= \argmin_{\xi,\theta,v,H,\sigma}  \ \mathbf{m}^T \mathbf{W} \mathbf{m} .
	\label{eqn: Parameter estimation, GMM objective}
\end{align} 
To improve statistical efficiency, we employ a feasible GMM procedure in which the weighting matrix is updated iteratively using the estimated covariance structure of the moment conditions. Specifically, the weighting matrix at each stage is taken as the inverse of an estimate of the covariance matrix of the fitted moment vector $\mathbf{m}$ obtained from the preceding GMM step. 

The procedure is initialized with the identity weighting matrix, $\mathbf{W}_1 = \mathbf{I}$. The first-stage estimator is therefore obtained from
\begin{align*}
	\{\xi_1^*, \theta_1^*, v_1^*,H_1^*,\sigma_1^*\}
	= \argmin_{\xi,\theta,v,H,\sigma}  \ \sum_{q=1}^4 \sum_{n=1}^{n_q} (\rho(q,n) - \hat{\rho}(q,n))^2,
\end{align*}
which corresponds to an equally weighted least-squares fit of the moment conditions.
Using the first-stage estimates $\{\xi_1^*, \theta_1^*, v_1^*,H_1^*,\sigma_1^*\}$, we evaluate the fitted moment vector $\mathbf{m}_1^*$ and construct the second-stage weighting matrix as
\begin{align*}
	\mathbf{W}_{2} = (\mathbf{m}_1^*\mathbf{m}_1^{*T})^{-1}.
\end{align*} 
The updated weighting matrix $\mathbf{W}_{2}$, is then used in the optimization problem defined by Equation~\eqref{eqn: Parameter estimation, GMM objective} to obtain the second-stage feasible GMM estimator. In practice, two iterations are typically sufficient, as the marginal improvement from additional iterations is generally negligible.

To mitigate the risk of convergence to local minima, we adopt a multi-start optimization strategy at each GMM stage. Initial parameter values are selected from all combinations of
$\xi_0 \in \{-1,0,1\}$,
$\theta_0 \in \{-1,0,1\}$,
$v_0 \in \{10^{-3},10^{-2},10^{-1}\}$, 
$H_0 \in \{0.3,0.5,0.7\}$, and 
$\sigma_0 \in \{0.05,0.1,0.2\}$. 
For each initialization, the optimization problem is solved independently, and the parameter estimates corresponding to the smallest value of the GMM objective function are retained. The weighting matrix used in subsequent iterations is then constructed from the fitted moment vector associated with this optimal solution.

\section{An Illustration using the S\&P500 Index}
\label{sec: Application to SP500}

To illustrate the fVG model and the proposed GMM estimation procedure, we use daily observations of the Standard \& Poor's 500 Index (S\&P 500) obtained from the CRSP database over a ten-year period spanning January 1, 2010 to December 31, 2019. For simplicity, observations are assumed to be equally spaced with sampling interval $d = 1/252$, corresponding to one trading day. The resulting dataset contains $N=2516$ daily price observations.

For the empirical analysis, we set the moments truncation to $n_1=n_2=n_3=n_4=p$ with $p \in \{2,3,4,5\}$, and implement the two-step feasible GMM procedure described in Section~\ref{sec: Parameter Estimation}. For any given value of $p$, the GMM framework incorporates $4p$ moment conditions. Increasing $p$ enables the moment conditions to capture the power-law temporal scaling more effectively, thereby improving identification and calibration of the Hurst exponent $H$. However, larger values of $p$ can also introduce numerical challenges, including multicollinearity among the moments, increased sampling variability, and a more ill-conditioned optimization problem, which can adversely affect estimation stability. Consequently, for empirical applications, we recommend using $p=2$, as it is the smallest value that achieves model identification, providing eight moment conditions for the estimation of five parameters.

In addition to estimating the full model, we consider several restricted specifications obtained by imposing the constraints $\theta=0$, $v=0$, and $H=0.5$ in all admissible combinations. For each restricted model, parameter estimation is carried out using the same GMM framework, with optimization performed over the corresponding constrained parameter space.
We exclude the specification with $v=0$ and $\theta\neq 0$. When $v=0$, the parameter $\theta$ is no longer identified because it enters the model only through the first moment in conjunction with $\xi$. Consequently, only the sum $\xi+\theta$ is identifiable. Intuitively, this specification introduces an additional deterministic drift term, $\theta t$, into the log-return process, which is already captured by existing the deterministic drift  $\xi t$.

Although more efficient estimation procedures are available for certain nested specifications, particularly when $H=0.5$, we adopt a common estimation methodology across all models to facilitate a direct comparison of parameter estimates. For example, the VG model admits likelihood-based estimation through its closed-form PDF. Nevertheless, employing a unified GMM framework ensures that differences in empirical performance can be attributed to the model specifications themselves rather than to differences in the estimation methodology.

\begin{table}[!htp]
	\centering 
	\resizebox{\ifdim\width>\linewidth\linewidth\else\width\fi}{!}{
	\begin{tabular}[t]{cccccccccc}
		\toprule
		\multicolumn{5}{c}{p = 2} & \multicolumn{5}{c}{p = 3} \\
		\cmidrule(l{3pt}r{3pt}){1-5} \cmidrule(l{3pt}r{3pt}){6-10}
		$\xi^*$ & $\sigma^*$ & $\theta^*$ & $v^*$ & $H^*$ & $\xi^*$ & $\sigma^*$ & $\theta^*$ & $v^*$ & $H^*$\\
		\midrule
		0.1048 & 0.1451 & - & - & - & 0.1048 & 0.1443 & - & - & -\\
		0.1048 & 0.1451 & - & 0.0099 & - & 0.1048 & 0.1443 & - & 0.0102 & -\\
		0.6932 & 0.1398 & -0.5885 & 0.0044 & - & 0.6471 & 0.1372 & -0.5422 & 0.0068 & -\\
		\addlinespace
		0.1048 & 0.1224 & - & - & 0.4659 & 0.1048 & 0.1300 & - & - & 0.4776\\
		0.1048 & 0.0828 & - & 0.0427 & 0.3491 & 0.1048 & 0.1301 & - & 0.0006 & 0.4776\\
		0.3481 & 0.1149 & -0.2433 & 0.0068 & 0.4511 & 1.0125 & 0.1178 & -0.9077 & 0.0037 & 0.4721\\
	\end{tabular}}
	\resizebox{\ifdim\width>\linewidth\linewidth\else\width\fi}{!}{
		\begin{tabular}[t]{cccccccccc}
			\toprule
			\multicolumn{5}{c}{p = 4} & \multicolumn{5}{c}{p = 5} \\
			\cmidrule(l{3pt}r{3pt}){1-5} \cmidrule(l{3pt}r{3pt}){6-10}
			$\xi^*$ & $\sigma^*$ & $\theta^*$ & $v^*$ & $H^*$ & $\xi^*$ & $\sigma^*$ & $\theta^*$ & $v^*$ & $H^*$\\
			\midrule
			0.1048 & 0.1431 & - & - & - & 0.1048 & 0.1420 & - & - & -\\
			0.1048 & 0.1430 & - & 0.0102 & - & 0.1048 & 0.1420 & - & 0.0100 & -\\
			0.6534 & 0.1344 & -0.5485 & 0.0080 & - & 0.4880 & 0.1350 & -0.3832 & 0.0132 & -\\
			\addlinespace
			0.1048 & 0.1266 & - & - & 0.4720 & 0.1048 & 0.1242 & - & - & 0.4679\\
			0.1048 & 0.0925 & - & 0.0849 & 0.3563 & 0.1048 & 0.1025 & - & 0.0441 & 0.4022\\
			0.6448 & 0.1113 & -0.5400 & 0.0084 & 0.4544 & 0.2898 & 0.1229 & -0.1851 & 0.0039 & 0.4652\\
			\bottomrule
	\end{tabular}}
	\caption{\label{tab:Estimated parameters of SP500}Estimated feasible GMM parameters for the S\&P 500 under alternative model specifications and moment truncation levels $n_1=n_2=n_3=n_4=p$.}
\end{table}

Table~\ref{tab:Estimated parameters of SP500} reports the optimal GMM estimates for all six combinations of parameter restrictions across moment truncation levels $p \in \{2,3,4,5\}$. Restricted parameters are indicated by a dash in the table.

We begin by discussing the main findings under our preferred specification, $p=2$.  
When the Hurst exponent is fixed at $H=0.5$, the estimated volatility parameter $\sigma^*$ is approximately $0.14$ across all model specifications. However, once $H$ is estimated freely, the fitted value of $\sigma^*$ declines. This pattern suggests that persistence captured by the Hurst exponent may otherwise be absorbed by the volatility parameter when temporal dependence is not adequately modeled. Stated differently, allowing $H$ to vary attributes part of the observed variability in S\&P 500 returns to long-range dependence rather than its instantaneous volatility.

When $\theta$ is fixed at zero, the estimated drift parameter $\xi^*$ is consistently around 0.1. Once $\theta$ is allowed to vary, however, the optimal value of $\xi^*$ increases because the first moment condition effectively identifies the combination $\xi^* + \theta^*$, which remains close to 0.16 across specifications.
The estimated $\theta^*$ is generally negative, indicating negative return skewness. This finding is consistent with the negative sample skewness of the S\&P500 reported by \citet{MadanCarrChang1998VGoption} for the VG model.

The estimated variance-intensity parameter $v^*$ is close to 0.01 when $\theta=0$ and $H=0.5$. Its value tends to increase when $H$ is estimated freely and decrease when $\theta$ is unrestricted. Since $v$ primarily governs the kurtosis of returns, this pattern suggests that erroneous omission of serial persistence $H$ leads to an underestimation of return kurtosis, whereas neglecting the skewness parameter $\theta$ causes excess kurtosis to absorb part of the asymmetry in the return distribution.

When $H$ is unrestricted, the estimated Hurst exponent generally lies within the interval $[0.35,0.5]$. 
Although these values remain relatively close to the Brownian benchmark $0.5$, the deviations are significant and economically meaningful. As noted above, even modest departures from $H \neq 0.5$ materially affect the estimation of the remaining parameters and imply substantially different stock-price dynamics.

Under the fractional BSM specification ($\theta=v=0$), the estimated Hurst exponent is typically around 0.47. Because higher-order moments vanish when $\theta=v=0$, estimation relies primarily on the first two moments, with $H$ entering through the second central moment. The results therefore suggest a slightly slower-than-linear decay in variance scaling over longer horizons relative to the Brownian benchmark.
Moreover, when kurtosis is explicitly modeled through the parameter $v$, the estimated Hurst exponent tends to decline further below 0.5 to 0.35. This finding indicates that part of the non-Gaussian heavy-tailed behavior of returns may otherwise be misattributed to temporal dependence, as both kurtosis and the Hurst exponent influence higher-order return moments. In contrast, under the fully unrestricted specification, the estimated Hurst exponent increases to approximately 0.45 once skewness and kurtosis are jointly accommodated through $\theta$ and $v$. 
Our result shows that the accurate calibration of $H$ requires that skewness and kurtosis are appropriately modeled.

Across different values of $p$, the qualitative findings remain largely unchanged. In particular, the parameter estimates, especially for $\xi^*$, $\sigma^*$ and $H^*$, are relatively stable within a given model specification. 
Nevertheless, estimation becomes less stable as $p$ increases beyond 5. For example, when $p=10$ (not shown), we occasionally obtain positive estimates of $\theta^*$ despite the negative empirical skewness of S\&P 500 returns. Parameter estimates often hit the boundaries and fluctuate erratically across initial starting values. 
Such results are consistent with the numerical difficulties discussed earlier and reinforce our preference for moderate truncation levels, particularly $p \in \{2,3\}$.

Interestingly, the estimated Hurst exponent is consistently below the Brownian benchmark of $H=0.5$. This finding suggests mild anti-persistence in S\&P 500 returns, whereby positive shocks are more likely to be followed by offsetting movements. Moreover, because the model separately accounts for skewness and excess kurtosis, the estimated Hurst exponent is less likely to absorb non-Gaussian features of the return distribution. Consequently, the results indicate that apparent long memory documented in simpler specifications may partly reflect unmodeled heavy tails rather than genuine temporal persistence.
Nonetheless, although $H<0.5$ is often interpreted as evidence of anti-persistent increments, the present GMM procedure identifies $H$ primarily through the scaling of return moments across investment horizons. Consequently, the estimates are more naturally interpreted as indicating that return moments accumulates slightly more slowly than the linear growth implied by standard Brownian motion. In other words, the term structure of empirical moments exhibits mildly sub-linear scaling, which the model captures through a Hurst exponent below 0.5.

A formal comparison of model performance remains challenging. The conditional distribution of the time-changed fractional Brownian motion does not admit a tractable closed-form expression, making the computation of forecasting metrics, likelihood-based criteria, and option-pricing implications considerably more demanding. A comprehensive evaluation of predictive and pricing performance is therefore left for future research.

\section{Conclusion}
\label{sec: Conclusion}

This paper introduces a novel framework for equity option pricing in which the stock prices are driven by a time-changed fBm. The proposed process preserves many of the key statistical properties that make fBm attractive for financial modeling, including its ability to capture long-range dependence, path roughness and anomalous diffusion. At the same time, the introduction of a stochastic time change ensures that time-changed fBm is a semimartingale process, thereby permitting arbitrage-free valuation within the standard framework of mathematical finance.

The proposed model provides a parsimonious yet rich framework for capturing salient features of observed asset-price dynamics. We hope that the results presented in this paper will stimulate further empirical and theoretical investigation, especially into the applications of asset pricing, risk management, and derivative valuation.

\appendix
\counterwithin{figure}{section}
\counterwithin{table}{section}
\printbibliography[title={References}]

\section{Proofs}
\label{sec: Proofs}

\subsection{Proof of Proposition \ref{prop: time changed fBm, unconditional raw moments}}
\label{ssec: Proof of Proposition: time changed fBm, unconditional raw moments}

For any $n \in \mathbb{N}$, we have
\begin{align*}
	\E \left[B^H(\gamma(t))^n \right]
	&= \E \left[ \E \left[ B^H(\gamma(t))^n | \gamma(t) \right] \right]
	= \begin{cases}
		\E \left[ \gamma(t)^{nH} (n-1)!!  \right] \\ 
		0 
	\end{cases} \hspace{-.1cm}
	= \begin{cases}
		v^{nH} \frac{\Gamma(t/v + nH)}{\Gamma(t/v)}  (n-1)!! , & n \text{ even} \\
		0, & n \text{ odd}
	\end{cases},
\end{align*}
where the second equality follows from the central moments of the Gaussian distribution and the third equality follows from the raw moments of the Gamma process.

\subsection{Proof of Theorem \ref{thm: time changed fBm, semimartingale X}}
\label{ssec: Proof of Theorem: X is a semimartingale}

Since the path of $B_{H}$ is continuous and $\gamma$ is a pure jump process, $X = B_{H}(\gamma)$ is also a pure jump process. 
In addition, $X$ is adapted and cadlag because $B_{H}$ and $\gamma$ are adapted and cadlag. 
Furthermore, by Proposition 2 of \citet{Yor2007GammaProcess}, $X$ has bounded variation on compacts. 
Thus, by Theorem 7 of \citet{Protter2005TBChapSemimartingales} (Chapter II, p. 55), $X$ is a semimartingale.

\subsection{Proof of Proposition \ref{prop: time changed fBm, Compensator of X}}

By Proposition 2 of \citet{Yor2007GammaProcess}, $X$ has bounded variation on compacts, and thus, the jumps of $X$ are bounded over $[0,T]$. 
It follows from Theorem 35 of \cite{Protter2005TBChapSemimartingales} (Chapter II, p.131) that $X$ is a special semimartingale.
Thus, by Theorem 34 of \cite{Protter2005TBChapSemimartingales} (Chapter II, p.130), $X$ admits a semimartingale decomposition $X = M+A$, where $X$ is a local martingale and $A$ is a unique, predictable, finite variation process. 

Next, we construct the compensator process $A = (A(t))_{t \in [0,T]}$. 
By the marked point process representation of $A$, we have  
{\allowdisplaybreaks
\begin{align*}
	A(t) 
	& =\int_{0}^{t}\int_{\mathbb{R}} x \, \nu_{X}(\diff s, \diff x)\\
	& =\int_{0}^{t}\int_{\mathbb{R}} x \, \psi_{X}(s,x) \, \diff x \, \diff s \\
	& =\int_{0}^{t}\int_{\mathbb{R}} x \left(\int_{0}^{\infty}\phi(x;\gamma(s),g) \, \psi_{\gamma}(g)\ \diff g\right) \, \diff x \, \diff s\\
	& =\int_{0}^{t}\int_{0}^{\infty}\left(\int_{\mathbb{R}} x\, \phi(x;\gamma(s),g) \, \diff x \right) \psi_{\gamma}(g)\, \diff g\, \diff s\\ 
	& =\int_{0}^{t}\int_{0}^{\infty}m_{B}(g;\gamma(s)) \, \psi_{\gamma}(g) \, \diff g \, \diff s 
\end{align*}}
for all $t \in [0,T]$. The existence of integrand also implies that $A$ is absolutely continuous, and thus, continuous.

Lastly, we show that $X-A$ is a true martingale. 
A sufficient condition is to show that $ \E[ (X(t)-A(t))^2] < \infty$ for all $t \in [0,T]$. 
By triangular/Minkowski's inequality
\begin{align*}
	\E\left[ (X(t)-A(t))^2\right]^{\frac{1}{2}} \leq  
	\E\left[ X(t)^2\right]^{\frac{1}{2}} 
	+ \E\left[ A(t)^2\right]^{\frac{1}{2}} ,
\end{align*}
it suffice to bound each term, $\E[ X(t)^2]$ and $\E[ A(t)^2]$, uniformly across $t \in [0,T]$. 
Firstly, using the iterated law of expectation and the fractional moments of the gamma distribution,
\begin{align*}
	\E[X(t)^2] 
	= \E[\E[B_H(\gamma(t)) | \gamma(t)]] 
	= \E[\gamma(t)^{2H}] 
	= \frac{\Gamma(t/v+2H)}{\Gamma(t/v)} v^{2H}
	< \infty
\end{align*}
for all  $t \in [0,T]$, $v > 0$ and $H \in (0,1)$.
Next, since $A$ is the predictable compensator of $X$, we know that $A(t) = \E[X(t) | \mathcal{F}(t-)]$. Thus, by Jensen's inequality,
\begin{align*}
	\E[A(t)^2] 
	= \E[\E[X(t) | \mathcal{F}(t-)]^2] 
	\leq \E[\E[X(t)^2 | \mathcal{F}(t-)]] 
	= \E[X(t)^2]
	< \infty
\end{align*}
for all  $t \in [0,T]$, $v > 0$ and $H \in (0,1)$.
This concludes the proof for elevating the local martingale into a true martingale.

\subsection{Proof of Proposition \ref{prop: fVG model, unconditional moments of W(t)}}
\label{ssec: Proof of Proposition: fVG model, unconditional moments of W(t)}

For any $t \geq 0$ and $h \geq 0 $, the increment of $W$ is 
\begin{align*}
	W(t+h) - W(t) = \theta G_h + \sigma G_h^H Z, 
\end{align*}
where $G_h = \gamma(t+h) - \gamma(t) \sim \Gamma(h/v, 1/v)$ and $Z \sim N(0,1)$ are independent. 
Then, the $n$-th raw moment of the increment $W(t+h) - W(t)$ is 
{\allowdisplaybreaks
\begin{align*}
	\E \left[ \left( W(t+h) - W(t) \right)^n \right]
	&= \E \left[  \left(\theta G_h + \sigma G_h^H Z \right)^n  \right] \\
	&= \E \left[ \sum_{j=0}^n  \binom{n}{j} (\theta G_h)^{n-j} \left(\sigma G_h^H Z \right)^j  \right] \\
	&= \sum_{j=0}^n  \binom{n}{j} \theta^{n-j}  \sigma^j \E \left[  G_h^{n-j(1-H)}  \right] \E \left[  Z^j  \right] \\
	&= \sum_{k=0}^{\lfloor n/2  \rfloor}  \binom{n}{2k} \theta^{n-2k}  \sigma^{2k} \E \left[  G_h^{n-2k(1-H)}  \right] \E \left[  Z^{2k}  \right] \\
	&= \sum_{k=0}^{\lfloor n/2  \rfloor}  \binom{n}{2k} \theta^{n-2k}  \sigma^{2k} 
	v^{n - 2k(1-H)} \frac{\Gamma \left(\frac{h}{v} + n-2k(1-H)\right)}{\Gamma \left(\frac{h}{v} \right)} (2k-1)!!  \\
	&= \sum_{k=0}^{\lfloor n/2  \rfloor} \frac{n!}{2^k k! (n-2k)!} \theta^{n-2k}  \sigma^{2k} 
	v^{n - 2k(1-H)} \frac{\Gamma \left(\frac{h}{v} + n-2k(1-H)\right)}{\Gamma \left(\frac{h}{v} \right)}.
\end{align*}}
Next, to establish the central moments, we evaluate the expectation
\begin{align*}
	\E[W(t+h) - W(t)]
	&= \theta \, \E\left[G_h \right]  
	= \theta h
\end{align*}
and express the centered increment as 
\begin{align*}
	W(t+h) - W(t) - \E[W(t+h) - W(t)] 
	= \theta (G_h - h) + \sigma G_h^H Z.
\end{align*} 
Consequently, for any $n \in \{2,3,...\}$, the $n$-th central moment is 
{\allowdisplaybreaks
\begin{align*}
	&\hspace{-.5cm}  \E \left[ \left( W(t+h) - W(t) - \E \left[   W(t+h) - W(t) \right] \right)^n \right] \nonumber \\ 
	&= \E \left[  \left(\theta (G_h-h) + \sigma G_h^H Z \right)^n  \right] \\
	&= \E \left[ \sum_{j=0}^n  \binom{n}{j} (\theta (G_h-h))^{n-j} \left(\sigma G_h^H Z \right)^j  \right] \\
	&= \sum_{j=0}^n  \binom{n}{j} \theta^{n-j}  \sigma^j \E \left[  (G_h-h)^{n-j} G_h^{jH}  \right] \E \left[  Z^j  \right] \\
	&= \sum_{k=0}^{\lfloor n/2  \rfloor}  \binom{n}{2k} \theta^{n-2k}  \sigma^{2k}  \E \left[  (G_h-h)^{n-2k} G_h^{2kH}  \right] \E \left[  Z^{2k}  \right] \\
	&= \sum_{k=0}^{\lfloor n/2  \rfloor}  \binom{n}{2k} \theta^{n-2k}  \sigma^{2k}  \sum_{\ell=0}^{n-2k} \binom{n-2k}{\ell} (-h)^{n-2k -\ell}  v^{2kH + \ell} \frac{\Gamma \left(\frac{h}{v} + 2kH + \ell \right)}{\Gamma \left(\frac{h}{v}\right)}  (2k-1)!! \\
	&= \sum_{k=0}^{\lfloor n/2  \rfloor}  \frac{n!}{2^k k!}   \theta^{n-2k}  \sigma^{2k}  \sum_{\ell=0}^{n-2k}  \frac{1}{\ell ! (n-2k-\ell)!} (-h)^{n-2k -\ell}  v^{2kH + \ell} \frac{\Gamma \left(\frac{h}{v} + 2kH + \ell \right)}{\Gamma \left(\frac{h}{v}\right)}    
\end{align*}}
since for any $k \in \{0,1,..., \lfloor n/2  \rfloor \}$, 
\begin{align*}
	\E \left[  (G_h-h)^{n-2k} G_h^{2kH}  \right]
	&= \E \left[ G_h^{2kH}  \sum_{\ell=0}^{n-2k} \binom{n-2k}{\ell}  (-h)^{n-2k -\ell} G_h^\ell  \right] \\
	&=   \sum_{\ell=0}^{n-2k} \binom{n-2k}{\ell} (-h)^{n-2k -\ell}  \E \left[G_h^{2kH + \ell}  \right] \\
	&=   \sum_{\ell=0}^{n-2k} \binom{n-2k}{\ell} (-h)^{n-2k -\ell}  v^{2kH + \ell} \frac{\Gamma \left(\frac{h}{v} + 2kH + \ell \right)}{\Gamma \left(\frac{h}{v}\right)} ,
\end{align*}  	
and the coefficients simplify to
\begin{align*}
	\binom{n}{2k} \binom{n-2k}{\ell} (2k-1)!! 
	= \frac{n!}{2^k k! (n-2k)!} \frac{(n-2k)!}{\ell ! (n-2k-\ell)!}
	= \frac{n!}{2^k k!}  \frac{1}{\ell ! (n-2k-\ell)!} .
\end{align*}

\subsection{Proof of Proposition \ref{prop: Equity model, Stock price dynamics under Q}}

We first recall the generalized Ito formula for semimartingale by \citet{Protter2005TBChapSemimartingales}. 
\begin{lemma}[Ito formula for semimartingale, Theorem 32 of \citet{Protter2005TBChapSemimartingales}]
	Let $Y$ be a semimartingale and $f \in C^{2}(\mathbb{R})$. 
	Then, the generalized Ito formula is
	\begin{align*}
		f(Y_t) &= f(Y_0) 
		+ \int_0^t \frac{\partial f}{\partial y} (Y_{s-}) \diff Y_s 
		+ \frac{1}{2}  \int_0^t \frac{\partial^2 f}{\partial y^2} (Y_{s-}) \diff [Y]_s^c \nonumber \\
		& \qquad + \sum_{s \in (0,t]} \left[f(Y_{s}) - f(Y_{s-}) - \frac{\partial f}{\partial y}(Y_{s-}) \Delta Y_{s} \right] . 
	\end{align*}
	\label{lemma: Ito formula for semimartingale}
\end{lemma}

Following the log returns process in Equation~\eqref{eqn: Equity model, fVG log stock returns process under P}, the discounted log returns process is given by 
\begin{align*}
	\diff \ln \bar{S}(t) = (\xi(t)  - r(t)) \diff t + \diff Y(t),
	\qquad \forall \, t \in [0,T].
\end{align*}
Applying the generalized Ito formula in Lemma~\ref{lemma: Ito formula for semimartingale} with $f(x)=e^x$ yields the discounted stock price process 
\begin{align*}
	\bar{S}(t) 
	=\bar{S}(0)
	+\int_{0}^{t} \bar{S}(c) \, (\xi(c) - r(c))  \diff c
	+\int_{0}^{t}\int_{\mathbb{R}} \bar{S}(c-) \left(e^{x}-1\right) \, \mu_Y(\diff c,\diff x) .
\end{align*}
Under $\mathbb{Q}$, the stock price process can be decomposed into its predictable drift and martingale components 
\begin{align*}
	\bar{S}(t) 
	&=\bar{S}(0)
	+\int_{0}^{t} \bar{S}(c) (\xi(c) - r(c)) \diff c  + \int_{0}^{t}\int_{\mathbb{R}}\bar{S}(c)\left(e^{x}-1\right)\tilde{\nu}_Y(\diff c,\diff x) \nonumber \\ 
	&\qquad\qquad +\int_{0}^{t}\int_{\mathbb{R}}\bar{S}(c-)\left(e^{x}-1\right)(\mu_Y-\tilde{\nu}_Y)(\diff c,\diff x) \nonumber \\
	&=\bar{S}(0)
	+\int_{0}^{t} \bar{S}(c) \left( \xi(c) - r(c) + \int_\mathbb{R} (e^x-1) \lambda(c,x) \psi_Y(c,x) \diff x \right) \diff c \nonumber \\ 
	&\qquad\qquad +\int_{0}^{t}\int_{\mathbb{R}}\bar{S}(c-)\left(e^{x}-1\right)(\mu_Y-\tilde{\nu}_Y)(\diff c,\diff x)
\end{align*}
where $\tilde{\nu}_Y(\diff t, \diff x) = \lambda(t,x)  \nu_Y(\diff t, \diff x)$.
Since $\bar{S}$ is a martingale under $\mathbb{Q}$, its predictable finite-variation component must vanish identically. Consequently,
\begin{align*}
	r(t) = \xi(t) + \int_\mathbb{R} (e^x-1) \, \lambda(t,x) \, \psi_Y(t,x) \, \diff x,
\end{align*}
which is precisely the arbitrage-free drift condition. Substituting this identity into the above decomposition yields Equation~\eqref{eqn: fVG model, Stock price under Q}.

\subsection{Proof of Proposition \ref{prop: Sufficient conditions for the existence of EMM}}

Under the strict positivity and Novikov conditions, Girsanov's theorem for marked point processes implies that the stochastic exponential
\begin{align*}
\frac{\diff \mathbb{Q}(\lambda)}{\diff \mathbb{P}}%
& =\exp\left\{ -\int_{0}^{T}\int_{\mathbb{R}}\left(\lambda(s,x)-1\right)\,\nu_{Y}(\diff s,\diff x)+\int_{0}^{T}\int_{\mathbb{R}}\log\lambda(s,x)\,\mu_{Y}(\diff s,\diff x)\right\}.
\end{align*} 
is a true $\mathbb{P}$-martingale. Consequently, it defines a probability measure $\mathbb{Q}(\lambda)$ equivalent to $\mathbb{P}$. 

Furthermore, under $\mathbb{Q}(\lambda)$, the compensator of the jump measure $\mu_Y$ is given by
\begin{align*}
	\tilde{\nu}_Y(\diff t, \diff x) = \lambda(t,x) \, \nu_Y(\diff t, \diff x).
\end{align*} 
Substituting this compensator into the stock price dynamics, the arbitrage-free drift condition ensures that the drift of the discounted stock price process vanishes. Finally, the true martingale condition guarantees that the discounted stock price is a true  $\mathbb{Q}(\lambda)$-martingale. Thus, $\mathbb{Q}(\lambda)$ is an EMM.

\end{document}